\documentclass[useAMS,usenatbib]{mnras}
\usepackage{multirow}
\usepackage{graphicx,color}
\newcommand{\etal }{{et al.} }

\newcommand{\vect}[1]{\mbox{\boldmath$#1$}}
\def\lesssim{\mathrel{\hbox{\rlap{\hbox{\lower4pt\hbox{$\sim$}}}\hbox{$<$}}}}
\def\gtrsim{\mathrel{\hbox{\rlap{\hbox{\lower4pt\hbox{$\sim$}}}\hbox{$>$}}}}
\newcommand{\cm}{\,{\rm cm}^{-3} } 
 
\newcommand{\nc}{n_{\rm c} }

\newcommand{\dfrac}[2]{{\displaystyle \frac{#1}{#2}} }

\newcommand{\zsun}{\thinspace Z_\odot}

\title[Evolution of Magnetic Fields]{Evolution of Magnetic Fields in Collapsing Star-forming Clouds under Different Environments}
\author[K. ~Higuchi, \etal]
{Koki Higuchi$^{1}$\thanks{E-mail: k.higuchi.521@s.kyushu-u.ac.jp (KH)}, Masahiro N. Machida$^{1}$ and Hajime Susa$^{2}$\\
$^{1}$ Department of Earth and Planetary Sciences, Faculty of Sciences, Kyushu University, Fukuoka 812-8581, Japan\\
$^{2}$Department of Physics, Konan University, Okamoto, Kobe 658-8501, Japan
}

\begin{document}
\maketitle
\begin{abstract}
In nearby star-forming clouds, amplification and dissipation of the magnetic field are known to play crucial roles in the star-formation process. 
The star-forming environment varies from  place to place and era to era in galaxies. 
In the present study, amplification and dissipation of magnetic fields in star-forming clouds are investigated under different environments using magnetohydrodynamics (MHD) simulations. 
We consider various star-forming environments in combination with the metallicity and the ionization strength, and prepare prestellar clouds having two different mass-to-flux ratios.
We calculate the cloud collapse until protostar formation using ideal and non-ideal (inclusion and exclusion of Ohmic dissipation and ambipolar diffusion) MHD calculations to investigate the evolution of the magnetic field.
We perform 288 runs in total and show the diversity of the density range within which the magnetic field effectively dissipates, depending on the environment.
In addition, the dominant dissipation process (Ohmic dissipation or ambipolar diffusion) is shown to strongly depend on the star-forming environment. 
Especially, for the primordial case, magnetic field rarely dissipates without ionization source, while it efficiently dissipates when very weak ionization sources exist in the surrounding environment. 
The results of the present study help to clarify star formation in various environments.
\end{abstract}

\section{Introduction}
\label{sec:intro}
Star formation has occurred in various locations of galaxies from the early universe to the present-day, and stars are a major component of the universe.
In addition, stars are closely related to galaxy formation and the chemical evolution of the universe.
Thus, the star formation process should be clarified in order to understand the history of the universe.

The star formation process in our galaxy has been investigated extensively through both observational and theoretical studies.
In particular, since low-mass star-forming regions are located in the neighborhood of the Sun, they have been observed in detail. 
Therefore, the star formation scenario is constructed based on observations of nearby star-forming regions \citep{andre93,kenyon95,jorgensen07, evans09, enoch09}. 
As a result, theoretical studies on present-day star formation are basically tuned to the environment of such star-forming regions \citep[e.g.,][]{shu87,mckee07,inutsuka12}.

There are two major problems in the star formation process \citep[the angular momentum and magnetic flux problmes,][]{mestel56,mestel85,mouschovias87,nakano72,basu94,mckee07}, namely, the angular momentum and magnetic flux in the star-forming clouds are many orders of magnitude (typically four to six) larger than those of protostars \citep{goodman93,caselli02,crutcher99,mckee07,crutcher10}. 
Thus, both the angular momentum and the magnetic flux should be removed from the clouds before star formation occurs. 
In the present-day (or nearby) star formation process, these problems are being investigated using three dimensional non-ideal magnetohydrodynamic (MHD) simulations, in which the angular momentum is transferred by magnetic braking and magnetically driven wind, and the magnetic field dissipates or is removed by both Ohmic dissipation and ambipolar diffusion \citep[e.g.,][]{tomisaka00,banerjee06, machida07,seifried11,tomida15,tsukamoto15,wurster16}. 
Thus, the magnetic field and its dissipation process are considered to be the keys to understanding present-day star formation.

The angular momentum and magnetic flux problems are the essence of star formation, and these problems are expected to exist in different star-forming environments of various galaxies during different eras. 
We need to investigate the evolution and dissipation of the magnetic field before studying the angular momentum transfer (or angular momentum problem), because the magnetic effects are closely related to the angular momentum transfer \citep{mouschovias79,nakano89,tomisaka00}.
Since the diffusivity of the magnetic field is determined primarily by the ionization degree (or the amount of charged particles) of star-forming clouds \citep[e.g.,][]{nakano02}, it should differ in each star-forming environment.
Note that the diffusivity also depends on the magnetic field strength \citep{wardle99}.
In order to estimate the magnetic diffusivity in different environments, \citet{susa15} investigated the dissipation of a magnetic field in various star-forming environments with different metallicities, ionization  (cosmic ray intensity and abundances of short- and long-lived radioactive elements) and magnetic field strengths using their one-zone model developed based on the model reported in \citet{omukai00}, \citet{omukai05} and \citet{omukai12} and clarified the region and density range in which magnetic dissipation becomes effective \cite[see also][]{maki04,maki07}. 

In addition to angular momentum transfer, the magnetic field plays an important role in the star formation process.
The magnetic field influences the final stellar mass and star formation efficiency \citep{matzner00}, which are closely related to the magnetically driven wind \citep{blandford82, uchida85}.
The formation of a circumstellar disk is related to the dissipation of the magnetic field  \citep{machida11b,li14,machida14a}.
In addition, the circumstellar disk is expected to evolve with the magneto-rotational instability \citep{balbus91,balbus98}.

Based on the aspect of the magnetic field and its dissipation,  \citet{susa15} speculated that the star formation process differs considerably in different environments. 
For example, the magnetic field rarely dissipates in primordial mini-halos and starburst galaxies with a relatively high ionization degree (or small magnetic diffusivity), while it significantly dissipates in our galaxy and more evolved galaxies with a lower ionization degree (or large magnetic diffusivity). 
Although using three-dimensional simulations, some studies focused on star formation in different environments with different metallicities \citep{jappsen07,jappsen09a,jappsen09b,dopcke11, dopcke13, bate14,chiaki16}, in addition to our previous studies \citep{machida08b, machida09, machida09b, machida15}, such studies ignored the effect and dissipation of magnetic field.
This is because the diffusion rate for the magnetic field, which is derived from the ionization degree of the star forming cloud, in different environments or different metallicities had not been investigated until they were estimated by \citet{susa15}. 
Unlike the case of calculating only the thermal evolution, many additional chemical networks are solved in order to estimate the magnetic diffusivities and ionization degree in collapsing star-forming clouds with different metallicities. 
Currently, the magnetic diffusivities in different environments can be estimated only by the one-zone calculation. 
On the other hand, the one-zone model can be used to discuss the dissipation/amplification of the magnetic field during gravitational collapse based only on order of magnitude arguments, because of the anisotropic nature of the Lorentz force. Hence, it is worth investigating the process through three-dimensional non-ideal MHD simulations.

In the present study, as the first step in investigating the star formation process at general locations in space and time, we calculate the evolution of star-forming clouds embedded in various environments considering the magnetic field, in which the thermal evolution and diffusion coefficients of Ohmic dissipation and ambipolar diffusion are taken from a table generated by one-zone calculations. 
In these calculations, as the initial condition, we adopt clouds with weak magnetic fields and low angular velocity in order to focus on the dissipation process of the magnetic field and compare the dissipation of the magnetic field derived from three-dimensional simulations with that derived from one-zone calculations (for details, see \S\ref{sec:methods} and \ref{sec:results}).

The remainder of the present paper is organized as follows. 
The numerical settings are described in \S\ref{sec:methods}, and the results are presented in \S\ref{sec:results}. We discuss the evolution and dissipation of magnetic fields in \S\ref{sec:discussion}. Finally, a summary is presented in \S\ref{sec:summary}.

\section{Methods}
\label{sec:methods}
\subsection{Parameters}
\label{sec:parameters}
We prepare 36 different star-forming environments, which are characterized by two parameters, the cloud metallicity $Z/\zsun$ and the ionization parameter $C_\zeta$. 
We consider the metallicity in the range of $0 \le Z/\zsun \le 1$ ($Z/\zsun=0$, $10^{-7}$, $10^{-6}$, $10^{-5}$, $10^{-4}$, $10^{-3}$, $10^{-2}$, $10^{-1}$, 1).
The ionization strength is defined as  
\begin{equation}
\zeta = \zeta_{\rm CR} + \zeta_{\rm RE, short} + \zeta_{\rm RE, long},
\label{eq:czeta}
\end{equation}
where the ionization rates associated with cosmic rays ($\zeta_{\rm CR}$), short-lived ($\zeta_{\rm RE, short}$) and long-lived ($\zeta_{\rm RE, long}$) radioactive elements are described as 
\begin{equation}
\zeta_{\rm CR} = C_{\zeta}\, \zeta_{\rm CR, 0}\, \rm{exp}(-\rho R_{\rm J} / \lambda),
\label{eq:cr}
\end{equation}
\begin{equation}
\zeta_{\rm RE, short} = 7.6 \times 10^{-19} \rm{s}^{-1}\, C_{\zeta}, 
\label{eq:short}
\end{equation}
and 
\begin{equation}
\zeta_{\rm RE, long} = 1.4 \times 10^{-22} \rm{s}^{-1}\, Z/\zsun.
\label{eq:long}
\end{equation}
where $\zeta_{\rm CR, 0}$ ($=1 \times 10^{-17}$\,s$^{-1}$), $R_{\rm J}$ and $\lambda$ ($=96$\,g\,cm$^{-2}$) mean the cosmic-ray ionization rate in the local ISM, the Jeans length and attenuation length, respectively.
In eqs. (\ref{eq:cr}) and (\ref{eq:short}), $C_\zeta$ is used as a parameter representing the ionization intensity of $\zeta_{\rm CR}$ and $\zeta_{\rm RE, short}$.
Although the ionization parameter, $C_\zeta$, associated with cosmic rays and short-lived radioactive elements can be independently specified, we simply represent them as described in eqs. (\ref{eq:cr})-(\ref{eq:short}) in order not to increase the number of parameters. 
This is reasonable because it is expected that an environment having a strong cosmic ray intensity, such as a starburst galaxy, has a large abundance of short-lived radioactive elements, which are ejected by supernovae.  
In addition, we assume that the ionization rate associated with long-lived radioactive elements is proportional to the metallicity $Z/\zsun$ (eq.~[\ref{eq:long}]), because such elements pile up over the course of the evolution of galaxies \citep[for more details, see][]{susa15}. 
In the present study, we adopt $C_{\zeta}=0$, 0.01, 1, and 10. 
The $C_{\zeta}$ values roughly mimic the following star-forming environments:
\begin{enumerate}
\renewcommand{\labelenumi}{(\roman{enumi})}
\item $C_\zeta=0$: Purely primordial environment, no ionization source is included
\item $C_\zeta=0.01$: Low-metallicity environment, weak ionization sources exist
\item $C_\zeta=1$: Nearby star-forming environment, the ionization intensity is the same as in nearby star-forming regions,
\item $C_\zeta=10$: Starburst galaxy environment: many (or strong) ionization sources exist 
\end{enumerate}

\begin{figure*}
\includegraphics[scale=0.44]{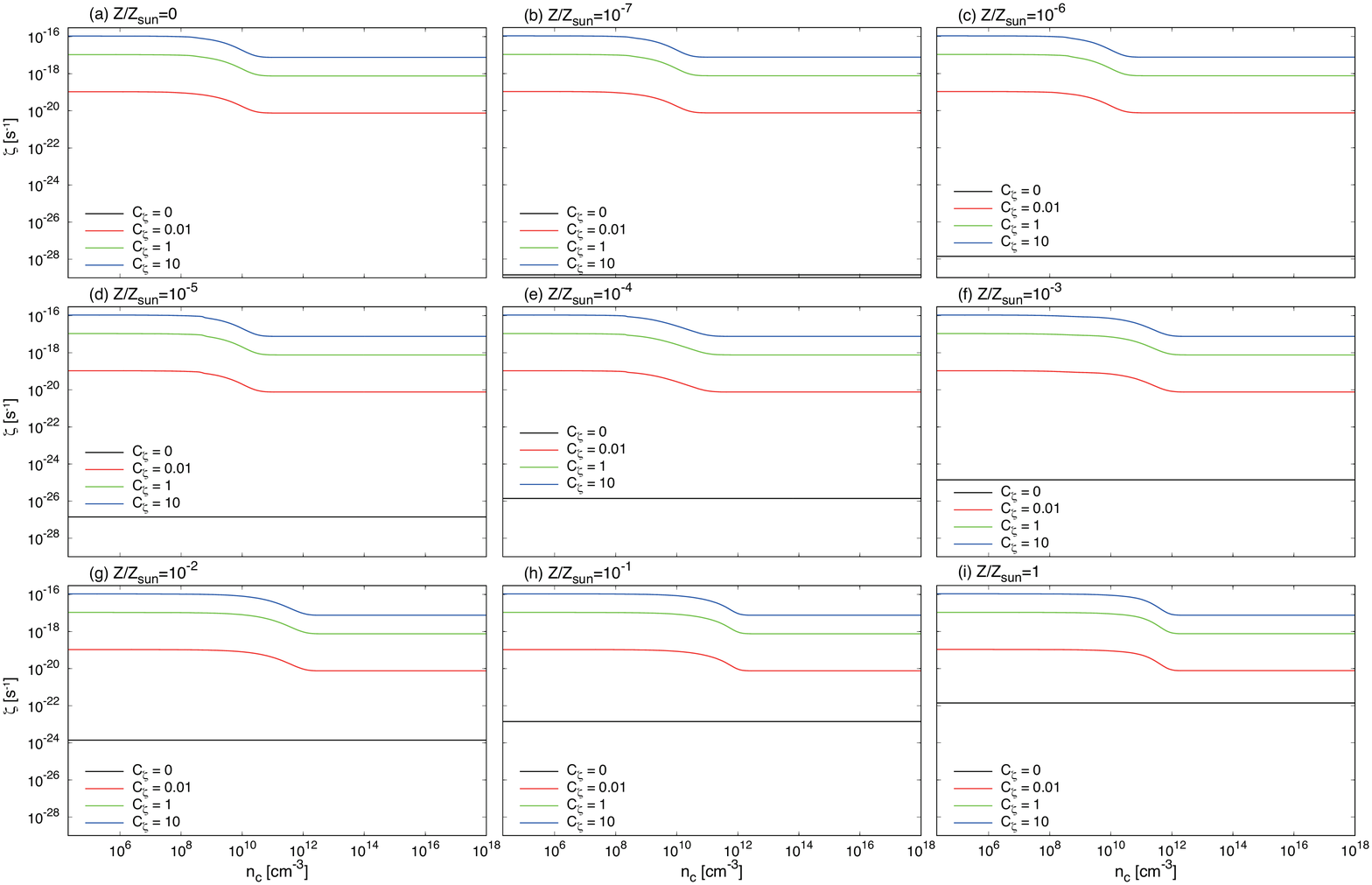}
\caption{
Ionization intensities of $C_\zeta$=0, 0.01, 1, and 10 for each metallicity against the central number density.
Parameters of $C_\zeta$ and $Z/\zsun$ are described in each panel.
}
\label{fig:zeta}
\end{figure*}

For reference, the ionization intensities of $C_\zeta$ = 0, 0.01, 1 and 10 for each metallicity ($Z/\zsun=0-1$) are plotted against the central number density in Fig.~\ref{fig:zeta}.

In addition to parameters $Z/\zsun$ and $C_{\zeta}$, two different magnetic field strengths are assumed in each environment (see \S 2.3). 
Combining these parameters, we investigate 72 prestellar (or star-forming) clouds embedded in different environments.
The model name and parameters $C_\zeta$ and $Z/\zsun$ are listed in Table~\ref{table:1}.

\begin{table*}
\centering
\caption{Model parameters.
Columns 1 and 2 list the model number and model name, respectively.
Columns 3 through 5 list the ionization strength $C_\zeta$, the metallicity $Z/\zsun$, and the mass-to-flux ratio $\mu_0$, respectively. 
Columns 6 through 10 list the initial magnetic field strength $B_0$, the initial angular velocity $\Omega_0$, the cloud mass $M_{\rm cl}$, the isothermal temperature $T_{\rm cl}$, and the cloud radius $r_{\rm cl}$, respectively.
}
    \scalebox{0.8}{
    \begin{tabular}{|c||c|c|c|c|c|c|c|c|c|c|c|} \hline
      &  Model & $C_{\zeta}$ & $Z/Z_{\rm{sun}}$ & $\mu_0$ & $B_0 (\rm{G})$ & $\Omega_0 (\rm{s^{-1}})$ & $M_{\rm{cl}} (M_{\odot})$ & $T_{\rm{cl}} (\rm{K})$ & $r_{\rm{cl}} (\rm{AU})$ \\ \hline

 1&      I0ZPM100 &     & $0$       &        & $1.02 \times 10^{-6}$ & $1.31 \times 10^{-17}$ & $1.08 \times 10^{4}$ & 198 & $4.91 \times 10^{5}$ &  \\
 2&      I0Z7M100 &     & $10^{-7}$ &        & $1.02 \times 10^{-6}$ & $1.31 \times 10^{-17}$ & $1.08 \times 10^{4}$ & 198 & $4.91 \times 10^{5}$ &  \\
 3&      I0Z6M100 &     & $10^{-6}$ &        & $1.02 \times 10^{-6}$ & $1.31 \times 10^{-17}$ & $1.07 \times 10^{4}$ & 198 & $4.91 \times 10^{5}$ &  \\
 4&      I0Z5M100 &     & $10^{-5}$ &        & $1.01 \times 10^{-6}$ & $1.31 \times 10^{-17}$ & $1.05 \times 10^{4}$ & 194 & $4.87 \times 10^{5}$ &  \\
 5&      I0Z4M100 & $0$ & $10^{-4}$ & $10^2$ & $9.55 \times 10^{-7}$ & $1.31 \times 10^{-17}$ & $8.75 \times 10^{3}$ & 172 & $4.59 \times 10^{5}$ &  \\
 6&      I0Z3M100 &     & $10^{-3}$ &        & $7.37 \times 10^{-7}$ & $1.31 \times 10^{-17}$ & $3.98 \times 10^{3}$ & 103 & $3.52 \times 10^{5}$ &  \\
 7&      I0Z2M100 &     & $10^{-2}$ &        & $2.95 \times 10^{-7}$ & $1.35 \times 10^{-17}$ & $2.27 \times 10^{2}$ & 16.4 & $1.33 \times 10^{5}$ & \\
 8&      I0Z1M100 &     & $10^{-1}$ &        & $3.09 \times 10^{-7}$ & $1.62 \times 10^{-17}$ & $1.26 \times 10^{2}$ & 18.1 & $9.67 \times 10^{4}$ & \\
 9&      I0Z0M100 &     & $1$       &        & $1.73 \times 10^{-7}$ & $1.78 \times 10^{-17}$ & $15.2$               & 5.65 & $4.49 \times 10^{4}$ &  \\ \hline
 10&      I0ZPM10000 &     & $0$       &        & $1.02 \times 10^{-8}$ & $1.31 \times 10^{-17}$ & $1.08 \times 10^{4}$ & 198 & $4.91 \times 10^{5}$ &  \\
 11&      I0Z7M10000 &     & $10^{-7}$ &        & $1.02 \times 10^{-8}$ & $1.31 \times 10^{-17}$ & $1.08 \times 10^{4}$ & 198 & $4.91 \times 10^{5}$ & \\
 12&      I0Z6M10000 &     & $10^{-6}$ &        & $1.02 \times 10^{-8}$ & $1.31 \times 10^{-17}$ & $1.07 \times 10^{4}$ & 198 & $4.91 \times 10^{5}$ &  \\
 13&      I0Z5M10000 &     & $10^{-5}$ &        & $1.01 \times 10^{-8}$ & $1.31 \times 10^{-17}$ & $1.05 \times 10^{4}$ & 194 & $4.87 \times 10^{5}$ &  \\
 14&      I0Z4M10000 & $0$ & $10^{-4}$ & $10^4$ & $9.55 \times 10^{-9}$ & $1.31 \times 10^{-17}$ & $8.75 \times 10^{3}$ & 172 & $4.59 \times 10^{5}$ &  \\ 
 15&      I0Z3M10000 &     & $10^{-3}$ &        & $7.37 \times 10^{-9}$ & $1.31 \times 10^{-17}$ & $3.98 \times 10^{3}$ & 103 & $3.52 \times 10^{5}$ &  \\
 16&      I0Z2M10000 &     & $10^{-2}$ &        & $2.95 \times 10^{-9}$ & $1.35 \times 10^{-17}$ & $2.27 \times 10^{2}$ & 16.4 & $1.33 \times 10^{5}$ &  \\
 17&      I0Z1M10000 &     & $10^{-1}$ &        & $3.09 \times 10^{-9}$ & $1.62 \times 10^{-17}$ & $1.26 \times 10^{2}$ & 18.1 & $9.67 \times 10^{4}$ &  \\
 18&      I0Z0M10000 &     & $1$       &        & $1.73 \times 10^{-9}$ & $1.78 \times 10^{-17}$ & $15.2$               & 5.65 & $4.49 \times 10^{4}$ &  \\ \hline
 19&      I001ZPM100 &        & $0$       &        & $8.51 \times 10^{-7}$ & $1.31 \times 10^{-17}$ & $6.20 \times 10^{3}$ & 137 & $4.09 \times 10^{5}$ &  \\
 20&      I001Z7M100 &        & $10^{-7}$ &        & $8.52 \times 10^{-7}$ & $1.31 \times 10^{-17}$ & $6.19 \times 10^{3}$ & 137 & $4.09 \times 10^{5}$ & \\
 21&      I001Z6M100 &        & $10^{-6}$ &        & $8.51 \times 10^{-7}$ & $1.31 \times 10^{-17}$ & $6.18 \times 10^{3}$ & 137 & $4.08 \times 10^{5}$ & \\
 22&      I001Z5M100 &        & $10^{-5}$ &        & $8.44 \times 10^{-7}$ & $1.31 \times 10^{-17}$ & $6.03 \times 10^{3}$ & 135 & $4.05 \times 10^{5}$ &  \\
 23&      I001Z4M100 & $0.01$ & $10^{-4}$ & $10^2$ & $7.87 \times 10^{-7}$ & $1.31 \times 10^{-17}$ & $4.88 \times 10^{3}$ & 117 & $3.77 \times 10^{5}$ &  \\
 24&      I001Z3M100 &        & $10^{-3}$ &        & $6.00 \times 10^{-7}$ & $1.31 \times 10^{-17}$ & $2.15 \times 10^{3}$ & 68.0 & $2.87 \times 10^{5}$ &  \\
 25&      I001Z2M100 &        & $10^{-2}$ &        & $2.95 \times 10^{-7}$ & $1.35 \times 10^{-17}$ & $2.30 \times 10^{2}$ & 16.5 & $1.34 \times 10^{5}$ &  \\
 26&      I001Z1M100 &        & $10^{-1}$ &        & $3.11 \times 10^{-7}$ & $1.62 \times 10^{-17}$ & $1.28 \times 10^{2}$ & 18.2 & $9.72 \times 10^{4}$ &  \\
 27&      I001Z0M100 &        & $1$       &        & $1.73 \times 10^{-7}$ & $1.78 \times 10^{-17}$ & $15.2$               & 5.64 & $4.49 \times 10^{4}$ & \\ \hline
 28&      I001ZPM10000 &        & $0$       &        & $8.51 \times 10^{-9}$ & $1.31 \times 10^{-17}$ & $6.20 \times 10^{3}$ & 137 & $4.09 \times 10^{5}$ &  \\
 29&      I001Z7M10000 &        & $10^{-7}$ &        & $8.52 \times 10^{-9}$ & $1.31 \times 10^{-17}$ & $6.19 \times 10^{3}$ & 137 & $4.09 \times 10^{5}$ &  \\
 30&      I001Z6M10000 &        & $10^{-6}$ &        & $8.51 \times 10^{-9}$ & $1.31 \times 10^{-17}$ & $6.18 \times 10^{3}$ & 137 & $4.08 \times 10^{5}$ &  \\
 31&      I001Z5M10000 &        & $10^{-5}$ &        & $8.44 \times 10^{-9}$ & $1.31 \times 10^{-17}$ & $6.03 \times 10^{3}$ & 135 & $4.05 \times 10^{5}$ &  \\
 32&      I001Z4M10000 & $0.01$ & $10^{-4}$ & $10^4$ & $7.87 \times 10^{-9}$ & $1.31 \times 10^{-17}$ & $4.88 \times 10^{3}$ & 117 & $3.77 \times 10^{5}$ &  \\
 33&      I001Z3M10000 &        & $10^{-3}$ &        & $6.00 \times 10^{-9}$ & $1.31 \times 10^{-17}$ & $2.15 \times 10^{3}$ & 68.0 & $2.87 \times 10^{5}$ &  \\
 34&      I001Z2M10000 &        & $10^{-2}$ &        & $2.95 \times 10^{-9}$ & $1.35 \times 10^{-17}$ & $2.30 \times 10^{2}$ & 16.5 & $1.34 \times 10^{5}$ &  \\
 35&      I001Z1M10000 &        & $10^{-1}$ &        & $3.11 \times 10^{-9}$ & $1.62 \times 10^{-17}$ & $1.28 \times 10^{2}$ & 18.2 & $9.72 \times 10^{4}$ &  \\
 36&      I001Z0M10000 &        & $1$       &        & $1.73 \times 10^{-9}$ & $1.78 \times 10^{-17}$ & $15.2$               & 5.64 & $4.49 \times 10^{4}$ &  \\ \hline
 37&      I1ZPM10000 &     & $0$       &        & $3.63 \times 10^{-9}$ & $1.31 \times 10^{-17}$ & $4.79 \times 10^2$ & 24.9 & $1.74 \times 10^{5}$ &  \\
 38&      I1Z7M10000 &     & $10^{-7}$ &        & $3.63 \times 10^{-9}$ & $1.31 \times 10^{-17}$ & $4.79 \times 10^2$ & 24.9 & $1.74 \times 10^{5}$ & \\
 39&      I1Z6M10000 &     & $10^{-6}$ &        & $3.63 \times 10^{-9}$ & $1.31 \times 10^{-17}$ & $4.79 \times 10^2$ & 24.9 & $1.74 \times 10^{5}$ &  \\
 40&      I1Z5M10000 &     & $10^{-5}$ &        & $3.64 \times 10^{-9}$ & $1.31 \times 10^{-17}$ & $4.82 \times 10^2$ & 25.1 & $1.74 \times 10^{5}$ &  \\
 41&      I1Z4M10000 & $1$ & $10^{-4}$ & $10^4$ & $3.71 \times 10^{-9}$ & $1.31 \times 10^{-17}$ & $5.09 \times 10^2$ & 26.0 & $1.77 \times 10^{5}$ &  \\
 42&      I1Z3M10000 &     & $10^{-3}$ &        & $3.80 \times 10^{-9}$ & $1.31 \times 10^{-17}$ & $5.43 \times 10^2$ & 27.3 & $1.81 \times 10^{5}$ &  \\
 43&      I1Z2M10000 &     & $10^{-2}$ &        & $3.64 \times 10^{-9}$ & $1.34 \times 10^{-17}$ & $4.39 \times 10^2$ & 25.0 & $1.66 \times 10^{5}$ &  \\
 44&      I1Z1M10000 &     & $10^{-1}$ &        & $3.26 \times 10^{-9}$ & $1.59 \times 10^{-17}$ & $1.58 \times 10^2$ & 20.1 & $1.06 \times 10^{5}$ &  \\
 45&      I1Z0M10000 &     & $1$       &        & $1.83 \times 10^{-9}$ & $1.78 \times 10^{-17}$ & $18.0$             & 6.34 & $4.75 \times 10^{4}$ &  \\ \hline
 46&      I1ZPM100 &     & $0$       &        & $3.63 \times 10^{-7}$ & $1.31 \times 10^{-17}$ & $4.79 \times 10^2$ & 24.9 & $1.74 \times 10^{5}$ &  \\
 47&      I1Z7M100 &     & $10^{-7}$ &        & $3.63 \times 10^{-7}$ & $1.31 \times 10^{-17}$ & $4.79 \times 10^2$ & 24.9 & $1.74 \times 10^{5}$ & \\
 48&      I1Z6M100 &     & $10^{-6}$ &        & $3.63 \times 10^{-7}$ & $1.31 \times 10^{-17}$ & $4.79 \times 10^2$ & 24.9 & $1.74 \times 10^{5}$ & \\
 49&      I1Z5M100 &     & $10^{-5}$ &        & $3.64 \times 10^{-7}$ & $1.31 \times 10^{-17}$ & $4.82 \times 10^2$ & 25.1 & $1.74 \times 10^{5}$ & \\
 50&      I1Z4M100 & $1$ & $10^{-4}$ & $10^2$ & $3.71 \times 10^{-7}$ & $1.31 \times 10^{-17}$ & $5.09 \times 10^2$ & 26.0 & $1.77 \times 10^{5}$ & \\
 51&      I1Z3M100 &     & $10^{-3}$ &        & $3.80 \times 10^{-7}$ & $1.31 \times 10^{-17}$ & $5.43 \times 10^2$ & 27.3 & $1.81 \times 10^{5}$ &  \\
 52&      I1Z2M100 &     & $10^{-2}$ &        & $3.64 \times 10^{-7}$ & $1.34 \times 10^{-17}$ & $4.39 \times 10^2$ & 25.0 & $1.66 \times 10^{5}$ &  \\
 53&      I1Z1M100 &     & $10^{-1}$ &        & $3.26 \times 10^{-7}$ & $1.59 \times 10^{-17}$ & $1.58 \times 10^2$ & 20.1 & $1.06 \times 10^{5}$ &  \\
 54&      I1Z0M100 &     & $1$       &        & $1.83 \times 10^{-7}$ & $1.78 \times 10^{-17}$ & $18.0$             & 6.34 & $4.75 \times 10^{4}$ &  \\ \hline
 55&      I10ZPM100 &      & $0$       &        & $4.05 \times 10^{-7}$ & $1.31 \times 10^{-17}$ & $6.56 \times 10^2$ & 31.0 & $1.93 \times 10^{5}$ &  \\
 56&      I10Z7M100 &      & $10^{-7}$ &        & $4.05 \times 10^{-7}$ & $1.31 \times 10^{-17}$ & $6.57 \times 10^2$ & 31.0 & $1.93 \times 10^{5}$ &  \\
 57&      I10Z6M100 &      & $10^{-6}$ &        & $4.05 \times 10^{-7}$ & $1.31 \times 10^{-17}$ & $6.57 \times 10^2$ & 31.0 & $1.93 \times 10^{5}$ &  \\
 58&      I10Z5M100 &      & $10^{-5}$ &        & $4.07 \times 10^{-7}$ & $1.31 \times 10^{-17}$ & $6.64 \times 10^2$ & 31.2 & $1.94 \times 10^{5}$ &  \\
 59&      I10Z4M100 & $10$ & $10^{-4}$ & $10^2$ & $4.19 \times 10^{-7}$ & $1.32 \times 10^{-17}$ & $7.25 \times 10^2$ & 33.1 & $1.99 \times 10^{5}$ &  \\
 60&      I10Z3M100 &      & $10^{-3}$ &        & $4.58 \times 10^{-7}$ & $1.32 \times 10^{-17}$ & $9.39 \times 10^2$ & 39.6 & $2.17 \times 10^{5}$ &  \\
 61&      I10Z2M100 &      & $10^{-2}$ &        & $4.58 \times 10^{-7}$ & $1.34 \times 10^{-17}$ & $8.67 \times 10^2$ & 39.6 & $2.09 \times 10^{5}$ &  \\
 62&      I10Z1M100 &      & $10^{-1}$ &        & $3.77 \times 10^{-7}$ & $1.55 \times 10^{-17}$ & $2.74 \times 10^2$ & 26.8 & $1.29 \times 10^{5}$ &  \\
 63&      I10Z0M100 &      & $1$       &        & $2.41 \times 10^{-7}$ & $1.78 \times 10^{-17}$ & $40.1$             & 11.0 & $6.24 \times 10^{4}$ &  \\ \hline
 64&      I10ZPM10000 &      & $0$       &        & $4.05 \times 10^{-9}$ & $1.31 \times 10^{-17}$ & $6.56 \times 10^2$ & 31.0 & $1.93 \times 10^{5}$ &  \\
 65&      I10Z7M10000 &      & $10^{-7}$ &        & $4.05 \times 10^{-9}$ & $1.31 \times 10^{-17}$ & $6.57 \times 10^2$ & 31.0 & $1.93 \times 10^{5}$ &  \\
 66&      I10Z6M10000 &      & $10^{-6}$ &        & $4.05 \times 10^{-9}$ & $1.31 \times 10^{-17}$ & $6.57 \times 10^2$ & 31.0 & $1.93 \times 10^{5}$ &  \\
 67&      I10Z5M10000 &      & $10^{-5}$ &        & $4.07 \times 10^{-9}$ & $1.31 \times 10^{-17}$ & $6.64 \times 10^2$ & 31.2 & $1.94 \times 10^{5}$ &  \\
 68&      I10Z4M10000 & $10$ & $10^{-4}$ & $10^4$ & $4.19 \times 10^{-9}$ & $1.32 \times 10^{-17}$ & $7.25 \times 10^2$ & 33.1 & $1.99 \times 10^{5}$ &  \\
 69&      I10Z3M10000 &      & $10^{-3}$ &        & $4.58 \times 10^{-9}$ & $1.32 \times 10^{-17}$ & $9.39 \times 10^2$ & 39.6 & $2.17 \times 10^{5}$ &  \\
 70&      I10Z2M10000 &      & $10^{-2}$ &        & $4.58 \times 10^{-9}$ & $1.34 \times 10^{-17}$ & $8.67 \times 10^2$ & 39.6 & $2.09 \times 10^{5}$ &  \\
 71&      I10Z1M10000 &      & $10^{-1}$ &        & $3.77 \times 10^{-9}$ & $1.55 \times 10^{-17}$ & $2.74 \times 10^2$ & 26.8 & $1.29 \times 10^{5}$ &  \\
 72&      I10Z0M10000 &      & $1$       &        & $2.41 \times 10^{-9}$ & $1.78 \times 10^{-17}$ & $40.1$             & 11.0 & $6.24 \times 10^{4}$ &  \\ \hline
    \end{tabular}
}
\label{table:1}
\end{table*}

\subsection{One-zone Model}
\label{sec:onezone}
We perform collapse simulations of a one-zone model in order to obtain the barotropic equation of state as well as the resistivity ($\eta_{\rm OD}$) and the ambipolar diffusion coefficient ($\eta_{\rm AD}$) for the three-dimensional MHD simulations. 
The model is equivalent to the model used in \cite{susa15}, in which the dynamics of the collapsing core is
 mimicked by the following equation:
\begin{equation}
\frac{d\rho}{dt}=\frac{\rho}{t_{\rm ff}}\sqrt{1-f},
\end{equation}
where the free-fall time $t_{\rm ff}$ is defined as
\begin{equation}
 t_{\rm ff} = \sqrt{\frac{3\pi}{32G\rho}},
\end{equation}
where $G$ is the gravitational constant, and $f$ is the ratio of the pressure gradient force to gravity at
the cloud centre \citep{omukai05}, and described as
\begin{equation}
f = \left\{
\begin{array}{l}
0,~~~~~~~~~~~~~~~~~~~~~~~~~~~~~~~~~~~~~~~~~~~~~~~ \gamma < 0.83, \\
0.6 + 2.5(\gamma - 1) - 6.0(\gamma - 1)^2,~~~~~~~~ 0.83 < \gamma < 1, \\
1.0 + 0.2(\gamma - 4/3) - 2.9(\gamma - 4/3)^2,~~~ \gamma > 1,
\end{array}
\right.\label{eq_f}
\end{equation}
where $\gamma$ ($\equiv d \rm{ln} \it{P} / \it{d} \rm{ln} \rho$) is the effective ratio of specific heat.
We also integrate the following energy equation for  
internal energy per unit mass $\epsilon$:
\begin{equation}
\frac{d\epsilon}{dt}=-p\frac{d}{dt}\left(\frac{1}{\rho}\right)
 -\Lambda_{\rm net} \label{eq_energy},
\end{equation}
where $\Lambda_{\rm net}$ is the net cooling rate for the gas, which 
includes the radiative cooling associated with [CII], [CI], [OI], H$_2$, HD, CO, OH, H$_2$O lines, continuum from the primordial gas and dust, cooling and heating associated with chemical reactions, 
the heating due to ionization by cosmic rays and radioactivity. 
In addition, the ordinary equation of state for an ideal gas is assumed:
\begin{equation}
p=(\gamma_{\rm ad} -1)\rho \epsilon,
\end{equation}
where $\gamma_{\rm ad}$ is the adiabatic index.

\begin{figure*}
\includegraphics[scale=0.60]{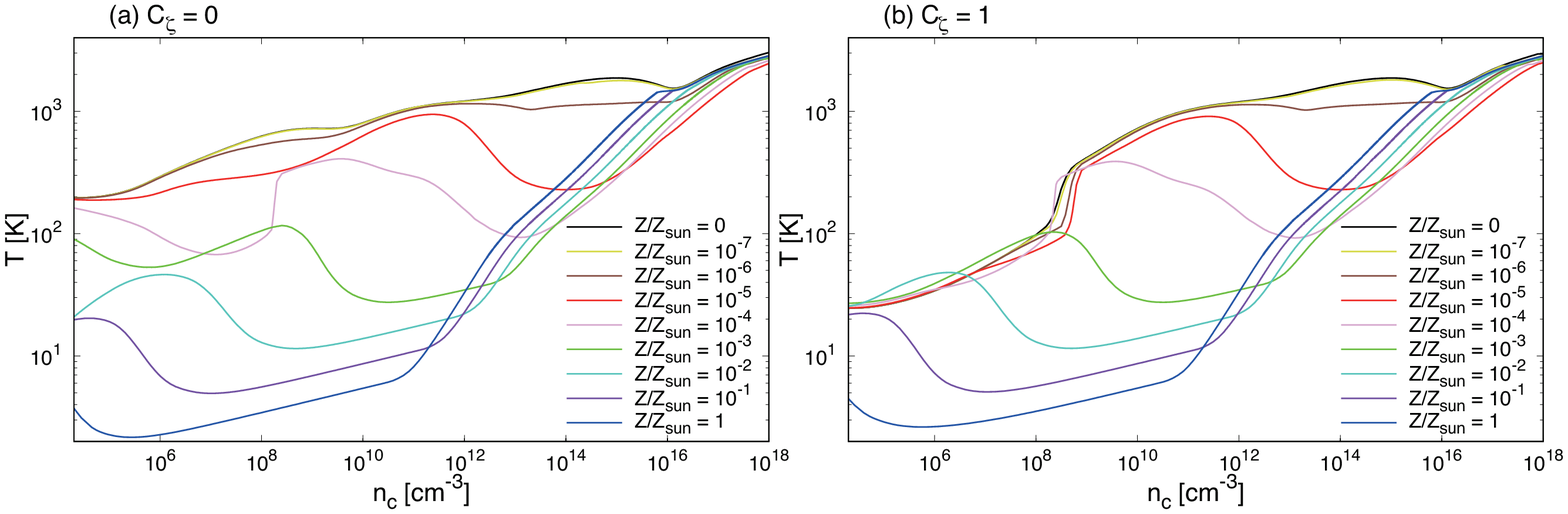}
\caption{
Temperatures with different metallicities ($Z/\zsun= 0-1$) given by one-zone calculations are plotted with respect to the central density for models with ({\it a}) $C_\zeta=0$ and ({\it b}) $1$. 
}
\label{fig:Temp}
\end{figure*}

We follow the fraction of the 59 chemical species and 5 charged/neutral dust
 grains by solving 
the non-equilibrium chemical network of 778 reactions \citep{susa15}. The network involves
H,
H$_2$,
e$^-$,
H$^+$,
H$_2^+$,
H$_3^+$,
H$^-$,
He,
He$^+$,
He$^{++}$,
HeH$^+$,
D,
HD,
D$^+$,
HD$^+$,
D$^-$,
C,
C$_2$,
CH,
CH$_2$,
CH$_3$,
CH$_4$,
C$^+$,
C$_2^+$,
CH$^+$,
CH$_2^+$,
CH$_3^+$,
CH$_4^+$,
CH$_5^+$,
O,
O$_2$,
OH,
CO,
H$_2$O,
HCO,
O$_2$H,
CO$_2$,
H$_2$CO,
H$_2$O$_2$,
O$^+$,
O$_2^+$,
OH$^+$,
CO$^+$,
H$_2$O$^+$,
HCO$^+$,
O$_2$H$^+$,
H$_3$O$^+$,
H$_2$CO$^+$,
HCO$_2^+$,
H$_3$CO$^+$,
Li,
Li$^+$,
Li$^{2+}$,
Li$^{3+}$,
Li$^-$,
LiH,
LiH$^+$,
M,
M$^+$,
G,
G$^+$,
G$^-$,
G$^{2+}$,
G$^{2-}$,
where M and G stand for the metallic elements and grains, respectively.

We expanded the reaction network of \citet{omukai05}  by adding the rates
 associated with Li and its ions/molecules listed in \citet{bovino} in
 order to
 properly assess the electron abundance at very low metallicity \citep{maki04}.
The reactions related with M and M$^+$ are also added, taken from \citet{umebayashi90} and \citet{prasad_huntress}.
The number fraction of M is assumed to be $y_{\rm M}=1.68\times 10^{-7}$
 \citep{umebayashi90} at the solar metallicity, and proportional to the metallicity.

Mass fraction of the dust assumed to be
$0.939\times 10^{-2}$ below the water-ice evaporation temperature
\citep{pollack94}, and proportional to the metallicity. 
The size distribution is \citep{mathis77}: 
\begin{equation}
\frac{dn_{\rm gr}}{da} \propto \left\{
\begin{array}{l}
a^{-3.5}~~~5\times 10^{-3} {\rm \mu m} < a < 1 {\rm \mu m} \\
a^{-5.5}~~~~~~~ 1{\rm \mu m} < a < 5 {\rm \mu m}.
\end{array}
\right.\label{eq_size}
\end{equation}
The collision rates between grain-charged particles
and grain-grain are calculated by eqs.(3.1)-(3.5) of \citet{drain_sutin}, 
averaged over the size distribution.

The direct ionization rates by cosmic rays are same as those of
\citet{omukai12}, except for the rates of M and HCO, which are taken from \citet{umebayashi90}.
The indirect ionization rates for M and HCO, i.e. the photoionization by the radiation
associated with the direct ionization by cosmic rays,  are taken from UMIST2 \citep{UMIST2}.

 As a result of the one-zone calculations, we obtain the thermal/chemical evolution of clouds with different 
 metallicity ($Z/Z_\odot$) and ionization strength $C_\zeta$.
Fig. \ref{fig:Temp} shows the evolutionary track of a collapsing gas cloud on the density-temperature plane. The left-hand panel shows the case of no ionization sources, whereas the right-hand panel shows the case for an ionization rate at the level of the interstellar medium. Each curve corresponds to the case of a certain metallicity, ranging
 from $Z/\zsun=0$ to $Z/\zsun=1$. The temperatures are tabulated as a function
 of density for use in the three-dimensional simulations described below.
 
With the parameter of magnetic field strength, we derive $\eta_{\rm OD}$ and $\eta_{\rm AD}$  from the fractions of charged species, which are obtained in each one-zone calculation.
The derivation of $\eta_{\rm OD}$ and $\eta_{\rm AD}$ can be referred to \citet{nakano86}, \citet{umebayashi90}, \citet{wardle99} and \citet{nakano02}. 
These are also tabulated as functions of density and magnetic field strength for each model.

\subsection{Initial Conditions}
We prepare a star-forming cloud in each environment listed in Table~\ref{table:1}.
Each cloud has a critical Bonnor-Ebert density profile \citep{ebert55,bonnor56}, which is uniquely determined when the central density $n_{\rm c,0}$ and isothermal temperature $T_{\rm cl}$ are given. 
We set $n_{\rm c,0}=10^4\cm$ as the initial central density.
The isothermal temperature in each environment is given by the one-zone calculation (\S\ref{sec:onezone}) and is described in the ninth column of Table~\ref{table:1}. 
The cloud radius $r_{\rm cl}$ depends on the initial cloud temperature as described in Table~\ref{table:1}.
In order to promote cloud contraction, we increase the density by 1.8 times to the critical Bonnor-Ebert density profile \citep{machida13}.
The initial cloud mass for each model is also listed in Table~\ref{table:1}. 
Although the initial clouds have different radii and masses with different metallicities, the ratio of thermal to gravitational energy ($\alpha_0$), which significantly affects the cloud collapse \citep[e.g.][]{miyama84,tsuribe99a,tsuribe99b}, is the same for all models ($\alpha_0=0.47$). 

The initial cloud also has the same ratio of rotational to gravitational energy $\beta_0=1.84 \times 10^{-8}$ for all models.
The magnetic field strength differs in each model. 
A uniform magnetic field is imposed over the entire computational domain, while the strength of the magnetic field is adjusted such that $\mu_0=10^2$ and $10^4$, where $\mu_0$ is the mass-to-flux ratio of the initial cloud normalized by the critical value, 
\begin{equation}
\mu_0 = \frac{\left( M/\Phi \right)}{\left( M/\Phi \right)_{\rm cri}},
\end{equation}
where $M$ and $\Phi$ are the mass and magnetic flux of the initial cloud, and the critical value is described as 
\begin{eqnarray}
    \Bigl( \frac{M}{\Phi} \Bigr)_{\rm{cri}} = \frac{1}{2 \pi G^{1/2}}.
\end{eqnarray}
The direction of the magnetic field vector is parallel to the rotation vector ($z$-axis).
The angular velocity and magnetic field strength are considerably small and only slightly affect the dynamics of the gas cloud for all models. 
In the present study, we focus on the evolution and dissipation of the magnetic field. 
A large angular velocity and strong magnetic field cause various phenomena, such as fragmentation, outflow and jet driving, and pseudo-disk and circumstellar disk formation. 
Before studying such phenomena, we should clarify (or confirm) the evolution and dissipation of the magnetic field in detail. 
With small $\Omega_0$ and weak $B_0$, we can purely investigate the amplification and dissipation of the magnetic field in a collapsing cloud immersed in different star-forming environments.

\subsection{Numerical Method and Basic Equations}
\label{sec:method}
In order to investigate the star formation process, we need to cover a considerably large spatial scale from the prestellar cloud ($\sim10^4$\,AU) to the protostar ($\sim0.01$\,AU). 
In the present study, the nested grid method is used to spatially resolve both the prestellar cloud and the protostar \citep[for a detailed description of the nested grid, see][]{machida04,machida05a,machida07,machida08}. 
Using the nested grid code, we resolve the Jeans wavelength of at least 16 cells \citep{truelove97}. 
Each grid is composed of $(i,j,k)=(64,64,32)$ cells. 
Mirror symmetry is imposed on the $z=0$ plane.
First, the fifth grids ($l=1-5$) are prepared, and the initial cloud is embedded in the fifth grid level ($l=5$). 
The low-density ($n_{\rm ISM}\simeq 10^3\cm$) interstellar medium is placed outside the initial cloud. 
Thus, the interstellar medium is fulfilled in the region of $r_{\rm cl}<r<16\, r_{\rm cl}$ in the $l=1-5$ grids. 
Such a large region of interstellar space prevents artificial reflection of Alfv\'en wave at the computational boundary \citep{machida13}. 
Note that, in this study, since we adopt considerably weak magnetic fields which have slow Alfv$\acute{\rm e}$n velocities (\S\ref{sec:parameters}), we do not need to impose such a large interstellar space.  
We set the large interstellar space for our future studies, in which strong magnetic fields should be adopted. 
The cell width halves with each increment in grid level. 
Although the physical sizes of the grid and the cell differ in each model, 30 grid levels ($l=30$) are used at maximum, where the cell width $h(l=30)$ of the maximum grid  is $h(l=30)\ll0.01$\,AU. 
\footnote{
Since the initial grid size is determined based on the initial cloud radius, it differs in each model with different metallicities.
}
Thus, the protostar is sufficiently resolved. 
The resolution study are presented in \S\ref{sec:appendix}.

The following basic equations are implemented in the nested grid code: 
\begin{eqnarray}
& \dfrac{\partial \rho}{\partial t} + \nabla \cdot (\rho \vect{v}) = 0, & \\
& \rho \dfrac{\partial \vect{v}}{\partial t} + \rho (\vect{v} \cdot \nabla) \vect{v} = - \nabla P -\dfrac{1}{4 \pi} \vect{B} \times (\nabla \times \vect{B}) - \rho \nabla \phi, & \\
& \dfrac{\partial \vect{B}}{\partial t} = \nabla \times \left[ \vect{v} \times \vect{B} + \dfrac{\eta_{\rm{AD}}}{|\vect{B}|^2}[(\nabla \times \vect{B}) \times \vect{B}] \times \vect{B} - \eta_{\rm{OD}} (\nabla \times \vect{B}) \right], & 
\label{eq:induction} \\
&    \nabla^2 \phi = 4 \pi G \rho, &
\end{eqnarray}
where $\rho$, $\vect{v}$, $P$, $\vect{B}$, and $\phi$ denote the density, velocity, pressure, magnetic flux density, and gravitational potential, respectively. 
The gas pressure $P$ is taken from the table derived by the one-zone calculation (\S\ref{sec:onezone}).
The coefficients of Ohmic resistivity $\eta_{\rm OD}$ and ambipolar diffusion $\eta_{\rm AD}$ are also taken from the table.
Note that $\eta_{\rm OD}$ is referred from the table as the argument of the density, while $\eta_{\rm AD}$ is referred as the argument of both the density and magnetic field strength. 
Note also that the dependence of magnetic field strength on $\eta_{\rm AD}$ are included in the table (see, \S\ref{sec:onezone}). 
The same procedure can be found in simulations of present-day star formation \citep{tsukamoto15}.
Thus, the gas pressure (or temperature), ionization degree, and diffusion coefficients ($\eta_{\rm OD}$ and $\eta_{\rm AD}$) are consistently given. 
The thermal evolution, chemical abundances, and magnetic diffusivities are described in \citet{susa15}.

Using these settings, we calculate the evolution of clouds until the central density reaches $n_{\rm c}\gtrsim10^{18}\cm$. 
Note that, in some models, we stop the calculation when the central density reaches $n_c\sim10^{16}\cm$ because the coefficient of magnetic diffusion is considerably large and the time step becomes extremely small (\S\ref{sec:results}).

In order to investigate the effect of magnetic diffusion in collapsing clouds, we perform four different calculations in each cloud  listed in Table~\ref{table:1}, as follows:
\begin{enumerate}
\renewcommand{\labelenumiii}{\alph{enumiii}.}
\item Ideal MHD calculation (hereafter Ideal): not including non-ideal MHD terms ($\eta_{\rm AD}=0$ and $\eta_{\rm OD}=0$) in  Eq.~(\ref{eq:induction})
\item Ohmic dissipation calculation (hereafter OD): including only Ohmic resistivity terms ($\eta_{\rm AD}=0$ and $\eta_{\rm OD}\ne0 $) in Eq.~(\ref{eq:induction})
\item Ambipolar diffusion calculation (hereafter AD): including only ambipolar diffusion terms ($\eta_{\rm AD}\ne0$ and $\eta_{\rm OD}=0 $) in Eq.~(\ref{eq:induction})
\item Non-ideal MHD calculation (hereafter Nonideal): including both  Ohmic resistivity and  ambipolar diffusion terms ($\eta_{\rm AD}\ne0$ and $\eta_{\rm OD}\ne0$) in Eq.~(\ref{eq:induction})
\end{enumerate}
In total, we calculate the evolution of 288 different clouds ($36\times2\times4$, 36 different environments, two different magnetic field strengths, and four different treatments of non-ideal MHD terms).

\section{Results}
\label{sec:results}
\subsection{Case of Clouds with a Mass-to-flux Ratio of $\mu_0=10^2$}
\label{sec:mu100}
\subsubsection{Typical Models}
\begin{figure*}
\includegraphics[scale=0.75]{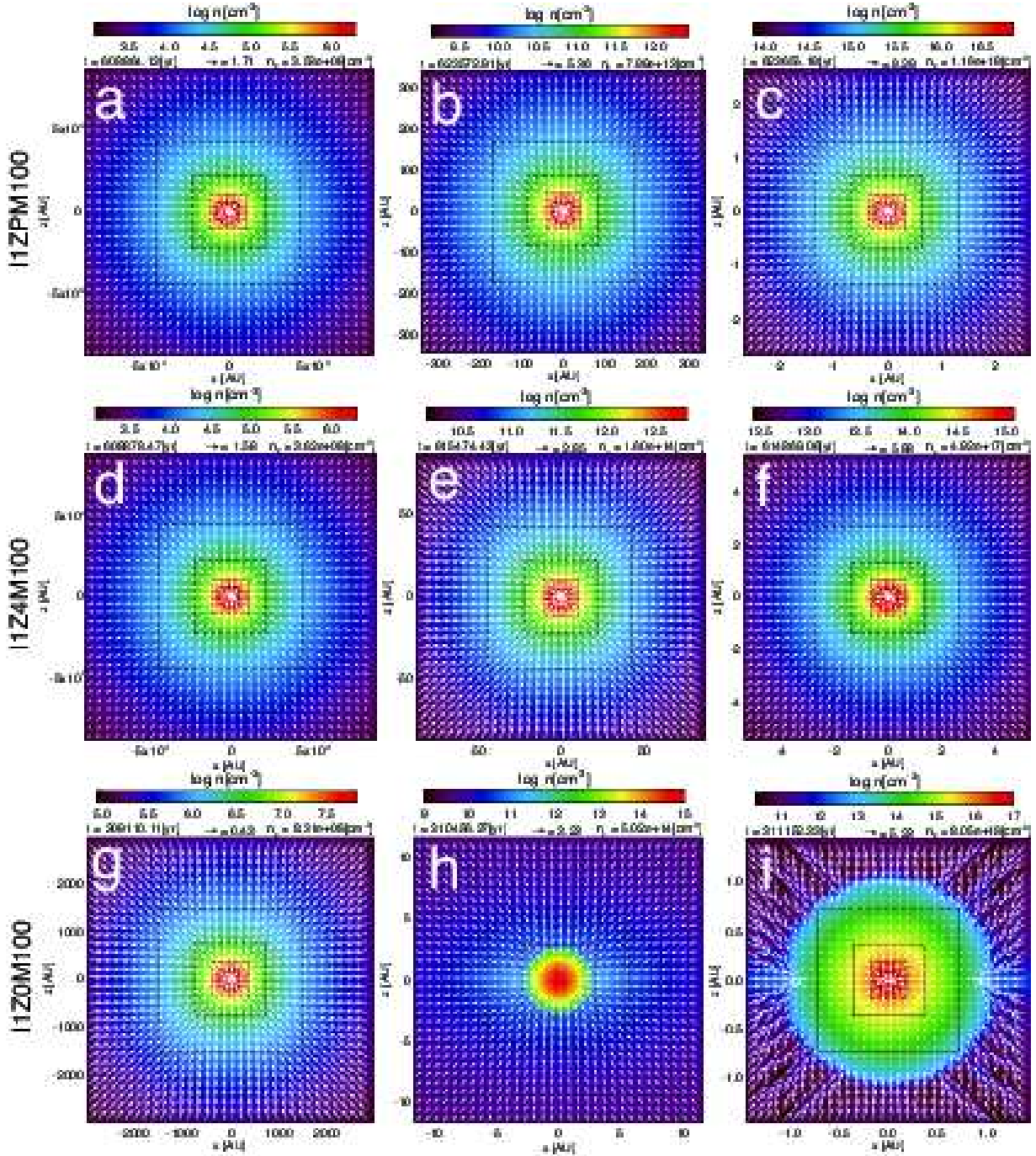}
\caption{
Time sequences of ({\it a}, {\it b}, {\it c}) models I1ZPM100, ({\it d}, {\it e}, {\it f}) I1Z4M100, and ({\it g}, {\it h}, {\it i}) I1Z0M100.
The density (colour) and velocity (arrows) distributions on the $y=0$ plane are plotted in each panel. 
The elapsed time and central number density are described in the upper part of each panel. 
The box scale differs in each panel.
The black squares in each panel indicate the boundary between grids.
}
\label{fig:mu100_time}
\end{figure*}

Fig.~\ref{fig:mu100_time} shows the time sequences for models I1ZPM100, I1Z4M100, and I1Z0M100.
The figure indicates that although the clouds become slightly oblate at higher density, the clouds collapse while maintaining nearly spherical symmetry.
The amplification of the magnetic field in the collapsing cloud depends on the collapse geometry \citep{scott80}. 
When the collapsing cloud maintains a nearly spherical symmetry, we can properly compare the dissipation rates of the magnetic field among models because the amplification rate of the magnetic field is approximately the same ($B \propto \rho^{2/3}$) without  magnetic dissipation \citep{machida07}. 
Fig.~\ref{fig:mu100_time} shows that each cloud collapses maintaining a nearly spherical symmetry at any epoch.
In Fig.~\ref{fig:mu100_time}{\it i}, we can see a shock front which corresponds to the first core. 
Note that, for model I1Z4M100, although the first core transiently appears, it disappears in a short time.

\begin{figure*}
\includegraphics[scale=0.55]{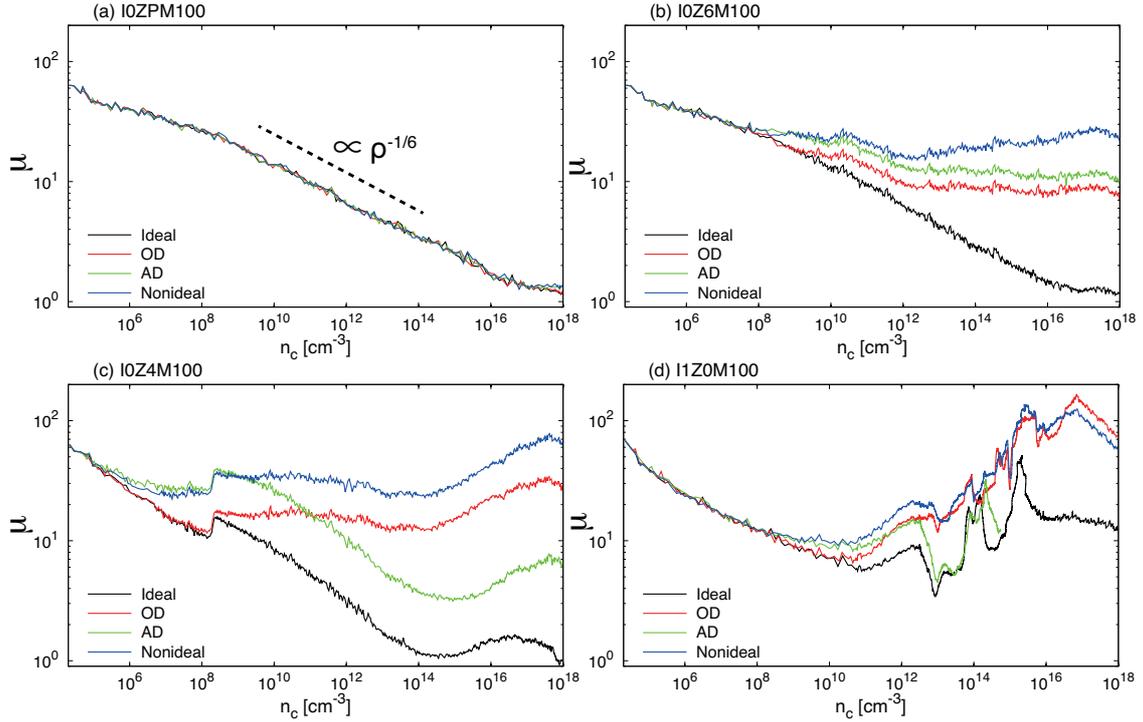}
\caption{
Mass-to-flux ratio $\mu$ with respect to the central number density $n_{\rm c}$ for ({\it a}) models I0ZPM100, ({\it b}) I0Z6M100, ({\it c}) I0Z4M100, and ({\it d}) I1Z0M100.
The $\mu$ of each calculation is estimated in the region of $\rho>0.1\rho_{\rm c}$, where $\rho_{\rm c}$ is the central density of each collapsing cloud.
The black, red, green, and blue solid lines, respectively, correspond to each calculation of Ideal, OD, AD, and Nonideal. 
A slope of $\propto \rho^{-1/6}$ is plotted in panel (a).
}
\label{fig:mu_each}
\end{figure*}

In order to confirm the dissipation of the magnetic field in collapsing clouds, the mass-to-flux ratios for typical models with  $\mu_0=10^2$ (I0ZPM100, I0Z6M100, I0Z4M100, and I1Z0M100) are plotted with respect to the central number density $n_{\rm c}$ in Fig.~\ref{fig:mu_each}. 
The mass-to-flux ratio $\mu$ at any epoch is estimated in the region of $n > 0.1 n_{\rm c}$. 
Thus, only the central region of the mass-to-flux ratio is plotted in the figure. 
We first show $C_{\zeta}=0$ models with metallicities $Z/\zsun=0$ (Fig.~\ref{fig:mu_each}{\it a}), $10^{-6}$ (Fig.~\ref{fig:mu_each}{\it b}), and $10^{-4}$ (Fig.~\ref{fig:mu_each}{\it c}), and the $C_{\zeta}=1$ model with metallicity $Z/\zsun=1$ (Fig.~\ref{fig:mu_each}{\it d}).

The first three models do not include an ionization source ($C_\zeta=0$), and the final model includes the same ionization intensity as in nearby star-forming regions ($C_\zeta=1$). 
Ionization sources are expected to be rare in the early universe ($Z/\zsun \sim 0$) or in the low-metallicity environment ($Z/\zsun \lesssim 10^{-4}$). 
However, nearby star-forming regions are expected to have a moderate ionization intensity ($C_\zeta=1$) and metallicity comparable to those for the Sun ($Z/\zsun=1$).
In the figure, the lines correspond to the calculation of ideal MHD (Ideal, black line), non-ideal MHD with only Ohmic dissipation (OD, red line), non-ideal MHD with only ambipolar diffusion (AD, green line), and non-ideal MHD with both Ohmic dissipation and ambipolar diffusion (Nonideal, blue line).

In Fig.~\ref{fig:mu_each}{\it a}, all calculations (Ideal, OD, AD, and Nonideal) track almost the same path as the mass-to-flux ratio. 
This indicates that non-ideal MHD effects are not significantly effective in the primordial environment ($C_\zeta=0$ and $Z/\zsun=0$), which is consistent with the findings of a previous study \citep{maki04}. 
The figure also indicates that the mass-to-flux ratio continues to decrease. 
Thus, the magnetic field continues to be amplified as the cloud collapses. 
The mass-to-flux ratio is approximately proportional to $\propto \rho^{-1/6}$ in Fig.~\ref{fig:mu_each}{\it a}. 
The mass-to-flux ratio is roughly described as $\mu \approx M /B L^2$, where the Jeans mass $M_{\rm J}$ and length $R_{\rm J}$ are adopted as the typical mass $M$ and length $L$.
Assuming $T \propto \rho^{\gamma -1}$, they are proportional to $ M_{\rm{J}} \propto \rho^{(3\gamma -4)/2} $ and $ R_{\rm{J}} \propto \rho^{(\gamma -2)/2}$.
With $B \propto \rho^{2/3}$, the mass-to-flux ratio is proportional to $\mu \propto \rho^{(3 \gamma -4) / 6}$. 
In the $Z/\zsun=0$ environment, $\gamma \approx 1-1.1$ \citep{omukai98,omukai00,omukai05,omukai10}. 
Adopting $\gamma=1$, we can derive $\mu \propto \rho^{-1/6}$.
The slight difference among models in Fig. \ref{fig:mu_each}{\it a} is attributed to the difference in non-ideal MHD effects, as described above.

The mass-to-flux ratios in Fig.~\ref{fig:mu_each}{\it b} begin to diverge at $n\sim10^8\cm$, indicating that non-ideal MHD effects become effective at this epoch.
\footnote{
We use the term of `diverge' when the deviation of the mass-to-flux ratio $\mu$ is seen among calculation models (Ideal, OD, AD and Nonideal models).
}
Thus, non-ideal MHD effects cannot be ignored in the nearly primordial environment ($C_\zeta=0$ and $Z/\zsun=10^{-6}$). 
The difference in the mass-to-flux ratio between Ideal and Nonideal calculations exceeds $10$ at $n_c=10^{18}\cm$. 
Since the magnetic field is inversely proportional to the mass-to-flux ratio, the magnetic field in the ideal MHD calculation (Ideal) is approximately 10 times stronger than that in the non-ideal MHD calculation (Nonideal). 
In addition, Fig.~\ref{fig:mu_each}{\it b} also indicates that the ambipolar diffusion rather than Ohmic dissipation greatly contributes to the magnetic dissipation. 
However, the Ohmic dissipation cannot be ignored because there is a significant difference for the Nonideal and OD calculations.
There is no significant difference in thermal evolution between the $Z/\zsun=0$ and $Z/\zsun=10^{-6}$ clouds, as shown in Fig.~1 of \citet{susa15}.
On the other hand, the difference in magnetic field between them is noticeable, which is attributed to the difference in the chemical abundance and ionization degree of the collapsing clouds \citep[for details, see][]{susa15}. 

The mass-to-flux ratio diverges at two different epochs ($n_{\rm c} \sim10^6\cm$ and $n_{\rm c} \sim10^8\cm$) for model I0Z4M100 (Fig.~\ref{fig:mu_each}{\it c}).
The figure indicates that ambipolar diffusion becomes effective at a lower density of $n_c\sim10^6\cm$, while Ohmic dissipation becomes effective at a higher density of  $n_c\sim10^8\cm$. 
Thus, the epochs at which ambipolar diffusion and Ohmic dissipation become effective clearly differ for this model. 
In addition, the difference in the magnetic field strength between Ideal and Nonideal calculations is approximately two orders of magnitude at the protostar formation epoch ($n_{\rm c} \sim 10^{18}\cm$). 
Therefore, the non-ideal MHD effects in this model are more effective than those in the I0ZPM100 and I0Z6M100 models.

The evolution of mass-to-flux ratios for model I1Z0M100 (Fig.~\ref{fig:mu_each}{\it d}), which has a moderate ionization intensity $C_\zeta=1$ and solar metallicity $Z/\zsun=1$, shows a similar tendency to models I0Z6M100 (Fig.~\ref{fig:mu_each}{\it b}) and I0Z4M100 (Fig.~\ref{fig:mu_each}{\it c}). 
Thus, even when a moderate ionization intensity exists, non-ideal MHD effects are important in the cloud with $Z/\zsun=1$, which has been extensively investigated in many previous studies \citep[e.g.,][]{nakano02}. 
In Fig.~\ref{fig:mu_each}{\it d}, strong oscillations of the mass-to-flux ratios at higher density ($n_c\gtrsim10^{12}\cm$) are attributed to the first core formation (see, Fig.~\ref{fig:mu100_time}{\it i}).
When the cloud has parameters of $C_\zeta=1$ and $Z/\zsun=1$, the first core is formed at $n_c\sim10^{11}\cm$ \citep{larson69,masunaga00}.
The first core is in a quasi-hydrostatic state after its formation.
Since the infalling gas remains around the first core, the gas continues to accrete to the first core. 
Thus, the self-gravity of the first core becomes strong with time, and the central density becomes high in order to recover to a hydrostatic state \citep{saigo06,saigo08}.  
Thus, the first core slowly contracts, oscillating around a hydrostatic state. 
The slow contraction and oscillation of the first core causes the oscillation of the magnetic field (or mass-to-flux ratio) inside the first core. 
One potential reason for the larger mass-to-flux ratio in model I1Z0M100, as compared to models I0Z6M100 and I0Z4M100, is the long lifetime of the first core for model I1Z0M100. 
The first core formation for different parameters is described in \S\ref{sec:discussion}.
It should be noted that the difference in mass-to-flux ratios between non-ideal and ideal MHD models for model I1Z0M100 is smaller than that for models I0Z6M100 and I0Z4M100, because model I1Z0M100 has a strong ionization intensity $C_\zeta=1$. 
When the ionization degree is high (or the ionization intensity is strong), the magnetic dissipation is not very effective.

In addition, Fig.~\ref{fig:mu_each} indicates that even when the initial cloud has the same mass-to-flux ratio, the amplification of the magnetic field strongly depends on the star-forming environment. 
Moreover, both ambipolar diffusion and Ohmic dissipation need to be included in order to better estimate the evolution of the magnetic field in collapsing clouds.

\begin{figure*}
\includegraphics[scale=0.55]{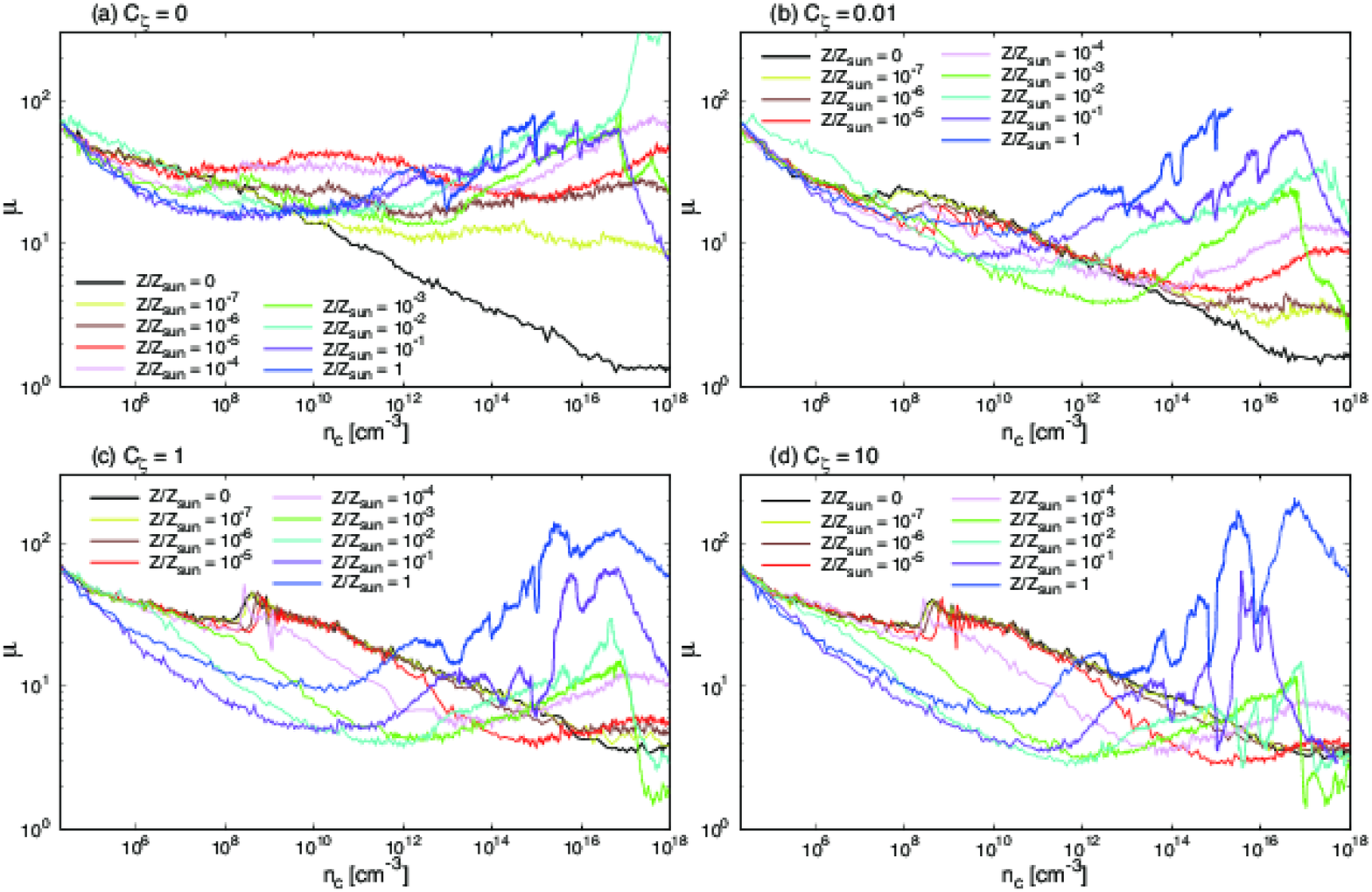}
\caption{
The mass-to-flux ratios for models with different metallicities ($Z/\zsun= 0-1$) are plotted with respect to the central density for different ionization strengths ({\it a}) $C_\zeta=0$, ({\it b}) $0.01$, ({\it c}) $1$, and ({\it d}) $10$. 
The $\mu$ of each model is estimated in the region of $\rho>0.1\rho_{\rm c}$, where $\rho_{\rm c}$ is the central density of each collapsing cloud.
Both Ohmic dissipation and ambipolar diffusion are included in the calculations.
}
\label{fig:mu_Czeta}
\end{figure*}
\subsubsection{Metallicity Dependence}
In order to investigate the dependence of the metallicity, the mass-to-flux ratios of collapsing clouds with different metallicities are plotted in each panel of Fig.~\ref{fig:mu_Czeta}, in which both Ohmic dissipation and ambipolar diffusion are included as the non-ideal MHD effects in the calculations.  
The figure indicates that the amplification of the magnetic field strongly depends on the metallicity  $Z/\zsun$.
In general, the magnetic fields in clouds with a lower metallicity tend to be more amplified than those in clouds with a higher metallicity. 
The coupling between the magnetic field (or ions and electrons) and neutrals is better in lower-metallicity clouds than in higher-metallicity clouds.
This is because low-metallicity clouds have a higher temperature and a small amount of dust grains \citep{susa15}.
The high temperature promotes the thermal ionization process, enriching the electrons and ions. 
As a result, the coupling between the magnetic field and neutrals is stronger via collisions with ions/electrons. 
In addition, the small amount of dust grains causes the high abundance of ions/electrons because the dust grains absorb these particles. 
As such, it is difficult for the charged dust particles to be the dominant charge carrier of electric current in low-metallicity gas, so that the conductivity is maintained relatively high. 
Thus, the range of density in which the Ohmic loss is effective becomes narrower as the metallicity decreases. 
In fact, the epoch at which the magnetic dissipation becomes effective clearly depends on the metallicity (see Figs.~\ref{fig:mu_Czeta}{\it b} through ~\ref{fig:mu_Czeta}{\it d}).
Note that the ionization rate associated with long-lived radioactive elements $\zeta_{\rm RE, long}$ is proportional to the metallicity $Z/\zsun$ (eq.~[\ref{eq:long}]) 
and thus it becomes low as the metallicity decreases as seen in Fig.~\ref{fig:zeta}. 
Nevertheless, the thermal ionization and deficit of dust grains cause a strong coupling between the magnetic field and neutrals even in lower metallicity clouds.

Fig.~\ref{fig:mu_Czeta} also indicates that even when the cloud initially has the same mass-to-flux ratio, the different amplification rates (or different dissipation rates) of the magnetic field in collapsing clouds with different metallicities produce differences in magnetic field strengths of one or two orders of magnitude at the protostar formation epoch. 
Note that the strong oscillations observed at higher density in models with higher metallicities (especially $Z/\zsun \ge 10^{-3}$) are related to the first core formation and are discussed in \S\ref{sec:discussion}.

\begin{figure*}
\includegraphics[scale=0.55]{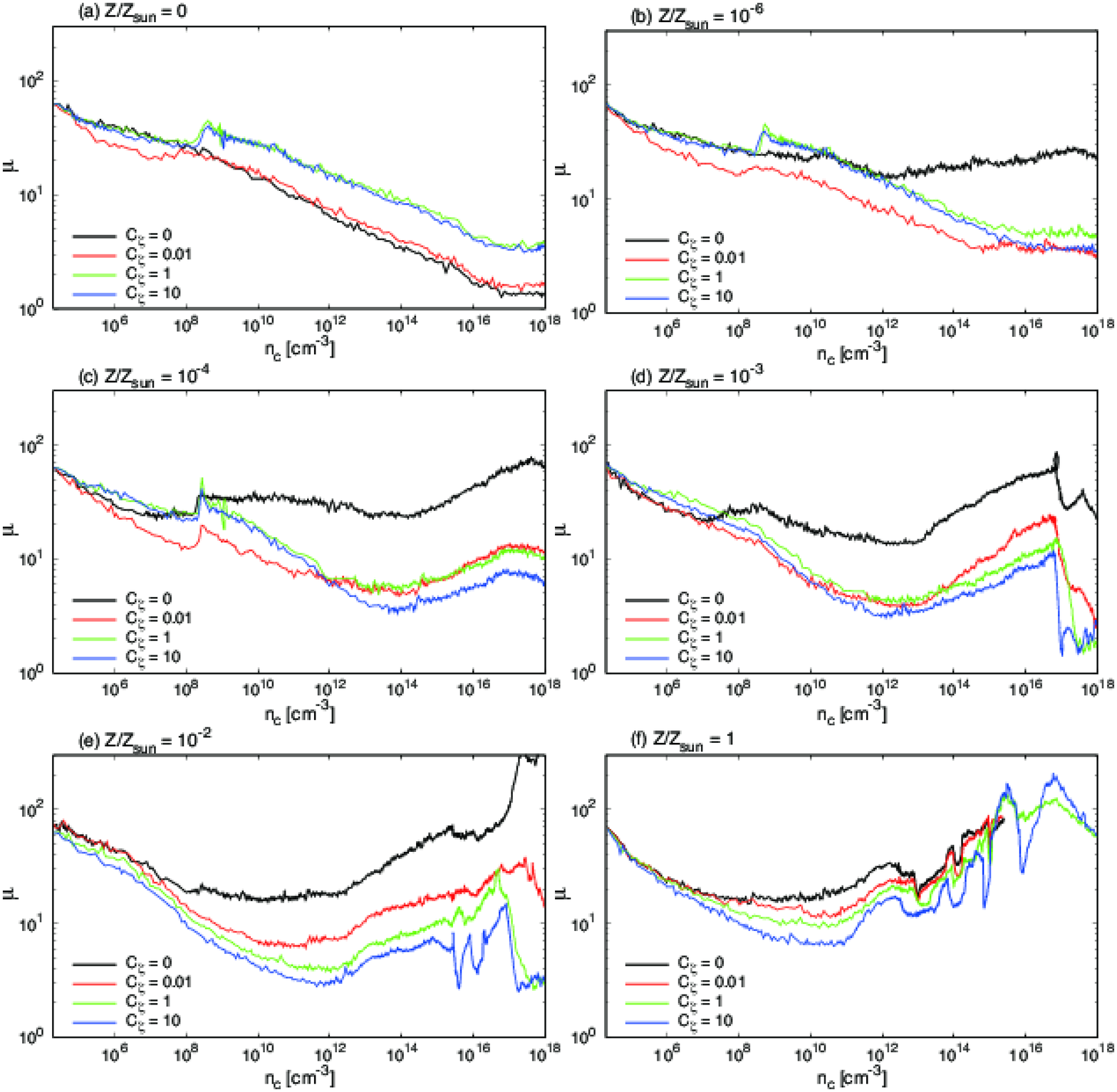}
\caption{
Mass-to-flux ratios for models with different parameters of ionization strength ($C_\zeta$=0, 0.01,1 and 10) are plotted against the central density for models with ({\it a}) $Z/\zsun=0$, ({\it b}) $10^{-6}$, ({\it c}) $10^{-4}$, ({\it d}) $10^{-3}$, ({\it e}) $10^{-2}$, and ({\it f}) 1. 
The $\mu$ of each model is estimated in the region of $\rho>0.1\rho_{\rm c}$, where $\rho_{\rm c}$ is the central density of each collapsing cloud.
Both Ohmic dissipation and ambipolar diffusion are included in the calculations.
}
\label{fig:mu_Z}
\end{figure*}
\subsubsection{Dependence on ionization Strength}
In order to investigate the effect of the ionization strength on the amplification of the magnetic field, mass-to-flux ratios for models with different $C_\zeta$ are plotted for different metallicities in Fig.~\ref{fig:mu_Z}, in which models with different  $C_\zeta$ are plotted in each panel.
Moreover, both Ohmic dissipation and ambipolar diffusion are included in the calculations.

The figure indicates that the magnetic field is further amplified and that protostars can have a strong magnetic field when the ionization intensity is large. This is natural because ionization sources of cosmic rays and short- and long-lived radioactive elements increase the ionization degree and promote coupling between the magnetic field and neutrals. 
Thus, in such environments, as the cloud collapses, the magnetic field is effectively amplified with inefficient magnetic dissipation. Note that, only for $Z/\zsun=0$ models (Fig. \ref{fig:mu_Z}{\it a}), stronger magnetic fields are realized in clouds with smaller $C_\zeta$ (or weaker ionization intensity), which is in contrast to the $Z/\zsun>0$ models.
This different trend in $Z/\zsun=0$ models is attributed to the quasi-static core formation at $n_c\sim10^8\cm$ (see \S\ref{sec:discussion}).

\begin{figure*}
\includegraphics[scale=0.55]{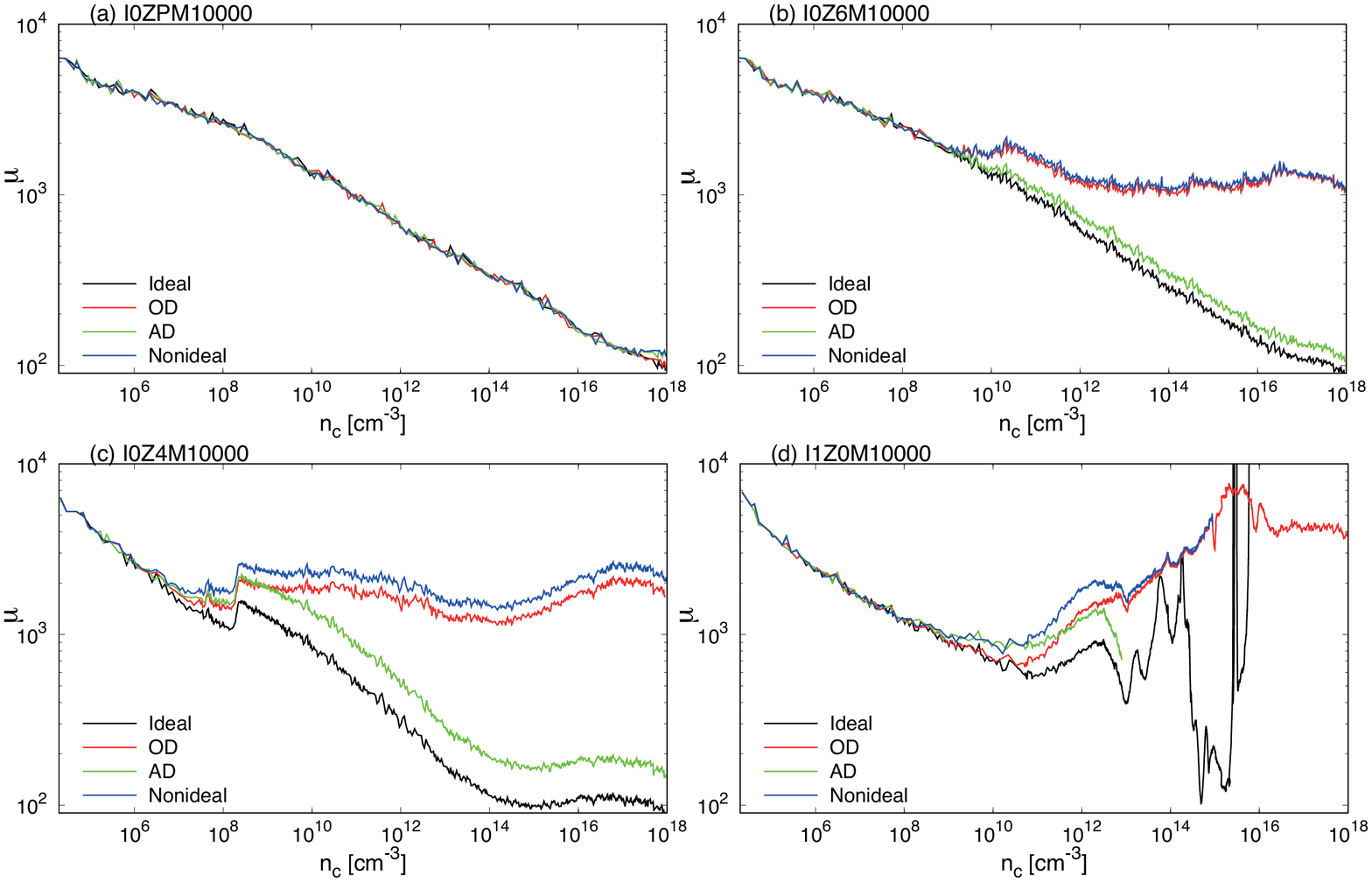}
\caption{
Same as Fig.~\ref{fig:mu_each}, but for $\mu_0=10^4$ models ({\it a}) I0ZPM10000, ({\it b}) I0Z6M10000, ({\it c}) I0Z4M10000, and ({\it d}) I1Z0M10000.
}
\label{fig:mu_eachterm}
\end{figure*}

\subsection{Case of Clouds with a Mass-to-flux Ratio of $\mu_0=10^4$}
The diffusivity of ambipolar diffusion depends on the magnetic field strength, whereas the Ohmic resistivity does not \citep{wardle99,nakano02}. 
In order to confirm the results described in \S\ref{sec:mu100}, we calculated the cloud evolution for the same environments as in \S\ref{sec:onezone} with a weaker magnetic field strength of $\mu_0=10^4$. 
Note that such weak fields ($\mu_0=10^2$ or $10^4$) are not unrealistic in the early universe or lower metallicity environments \citep{widrow02,ichiki06,xu08,ando10,doi11,widrow12,shiromoto14}. 
However, at the present day (or in our galaxy), the magnetic field is strong and the star-forming clouds have smaller mass-to-flux ratios of $\mu_0 \lesssim 10$ \citep{crutcher99,crutcher10}. 
In the present study, we adopted weak magnetic fields in order to compare the evolution of the magnetic field among models with the same condition and settings. 
In a forthcoming paper, we will investigate stronger magnetic field cases, in which we will focus on the outflow and jet driving and circumstellar disk formation. 

Fig.~\ref{fig:mu_eachterm} is the same as Fig.~\ref{fig:mu_each}, but for $\mu_0=10^4$ models.
Overall, the tendencies observed in the $\mu_0=10^2$ models (Fig.~\ref{fig:mu_each}) do not change in the $\mu_0=10^4$ models (Fig.~\ref{fig:mu_eachterm}). 
However, there exists a quantitative difference between the $\mu_0=10^2$ and $\mu_0=10^4$ models. 
For the primordial case ($Z/\zsun=0$ models; Figs.~\ref{fig:mu_each}{\it a} and \ref{fig:mu_eachterm}{\it a}),  there is no significant difference in the evolution track of $\mu$ among calculation models (Ideal, OD, AD and NonIdeal). 
In addition, the slope of $\mu$  in Fig.~\ref{fig:mu_eachterm}{\it a} is the same as that in Fig.~\ref{fig:mu_each}{\it a}. 

On the other hand, comparing Figs.~\ref{fig:mu_each}{\it b}--{\it d} with Figs.~\ref{fig:mu_eachterm}{\it b}--{\it d}, there exist quantitative differences in $Z/\zsun>0$ models. 
For example, the difference in the mass-to-flux ratio $\mu$ between AD and Ideal  is within a factor of  $\lesssim 1.5$ at $n_c \sim 10^{18}~{\rm cm^{-3}}$ for the $\mu_0=10^4$ models (Figs.~\ref{fig:mu_eachterm}{\it b} and {\it c}), whereas the difference in $\mu$ between AD and Ideal is as large as a factor of $\gtrsim 5.5$ at the same epoch for the $\mu_0=10^2$ models (Fig.~\ref{fig:mu_each}{\it b} and {\it c}). 
Thus, the difference in magnetic field between Ideal (black lines) and AD (green lines) in the $\mu_0=10^2$ models (Fig.~\ref{fig:mu_each}) is larger than that in the $\mu_0=10^4$ models (Fig.~\ref{fig:mu_eachterm}).
This indicates that the effect of ambipolar diffusion is not very strong in clouds with $\mu_0=10^4$. 
This is natural because the Lorentz force or the magnetic tension force, which detaches the magnetic field from neutrals, is weak in the $\mu_0 = 10^4$ models.

\begin{figure*}
\includegraphics[scale=0.48]{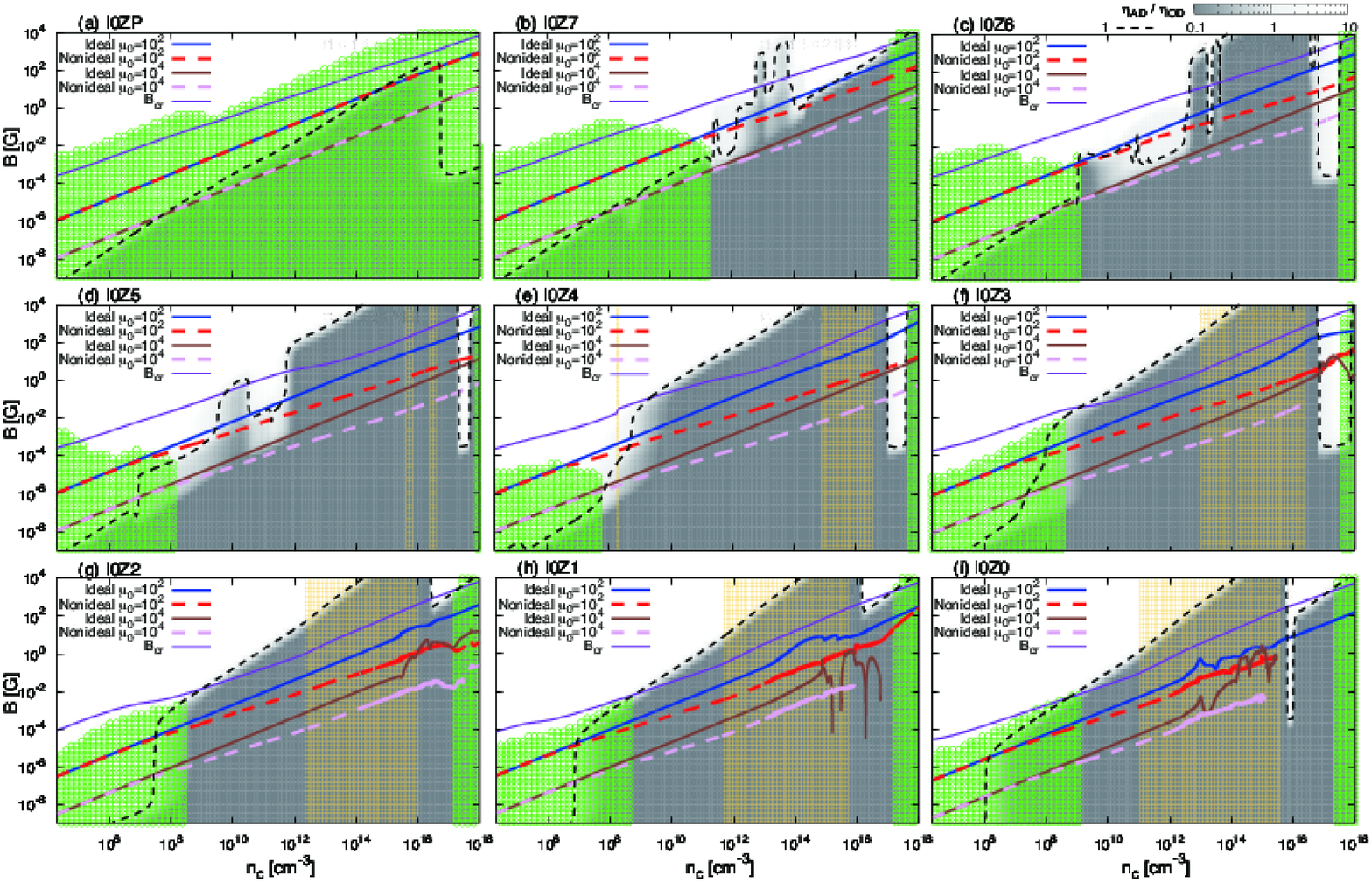}
\caption{
Evolution of the magnetic field strength $B$ at the cloud centre with respect to the central number density $n_{\rm c}$ for ideal and non-ideal MHD calculations; $\mu_0=10^2$ and $10^4$ calculation results are plotted in each panel. 
The critical magnetic field strength $B_{\rm cr}$ is also shown by the purple line.
The green region corresponds to the magnetically active region in which the magnetic field is well coupled with neutrals, whereas the magnetic field is not coupled with neutrals in other regions (grey and white regions). 
The grey colour represents the ratio of ambipolar diffusivity $\eta_{\rm AD}$ to Ohmic resistivity $\eta_{\rm OD}$.
The ratio of $\eta_{\rm AD}/\eta_{\rm OD}=1$ is plotted as the black broken line, above which ambipolar diffusion dominates Ohmic dissipation as the dissipation process of the magnetic field. 
The yellow hatched region corresponds to the adiabatic index exceeding $\gamma>4/3$ ($P\propto \rho^\gamma$), where a quasi-hydrostatic core may form.
In each panel, the metallicity parameter $Z/\zsun$ differs ($Z/\zsun=0$, $10^{-7}$, $10^{-6}$, $10^{-5}$, $10^{-4}$, $10^{-3}$, $10^{-2}$, $10^{-1}$ and $1$), but the ionization parameter $C_\zeta$ is the same ($C_\zeta=0$).
The model name is indicated in the upper left corner of each panel.
}
\label{fig:color0}
\end{figure*}
\section{Discussion}
\label{sec:discussion}
In this section, we describe the magnetic field dissipation on the
central density vs. B-field strength diagram, i.e., $n_{\rm c}-B$ plane,
which has often been used in the previous one-zone
approach \citep[e.g.,][]{nakano86}. Hence, this section also compares the three-dimensional MHD results with the assessment based on the
comparison of time scales in one-zone models.
In addition, we discuss the relation between the magnetic dissipation and
the formation of the adiabatic core (or the first core), which is important for
the formation of outflows, fragmentation, jets, etc.
We also discuss the possible implications of the formation of HMP stars (Hyper Metal-Poor stars, which have a metallicity of $Z/\zsun \le 10^{-5}$) and EMP stars (Extremely Metal-Poor stars, which have a metallicity $-4 < Z/\zsun < -3$; \citet{bc05}).

\subsection{Magnetic field dissipation on the $n_{\rm c}-B$ plane}
\label{sec:nBplane}
\begin{figure*}
\includegraphics[scale=0.48]{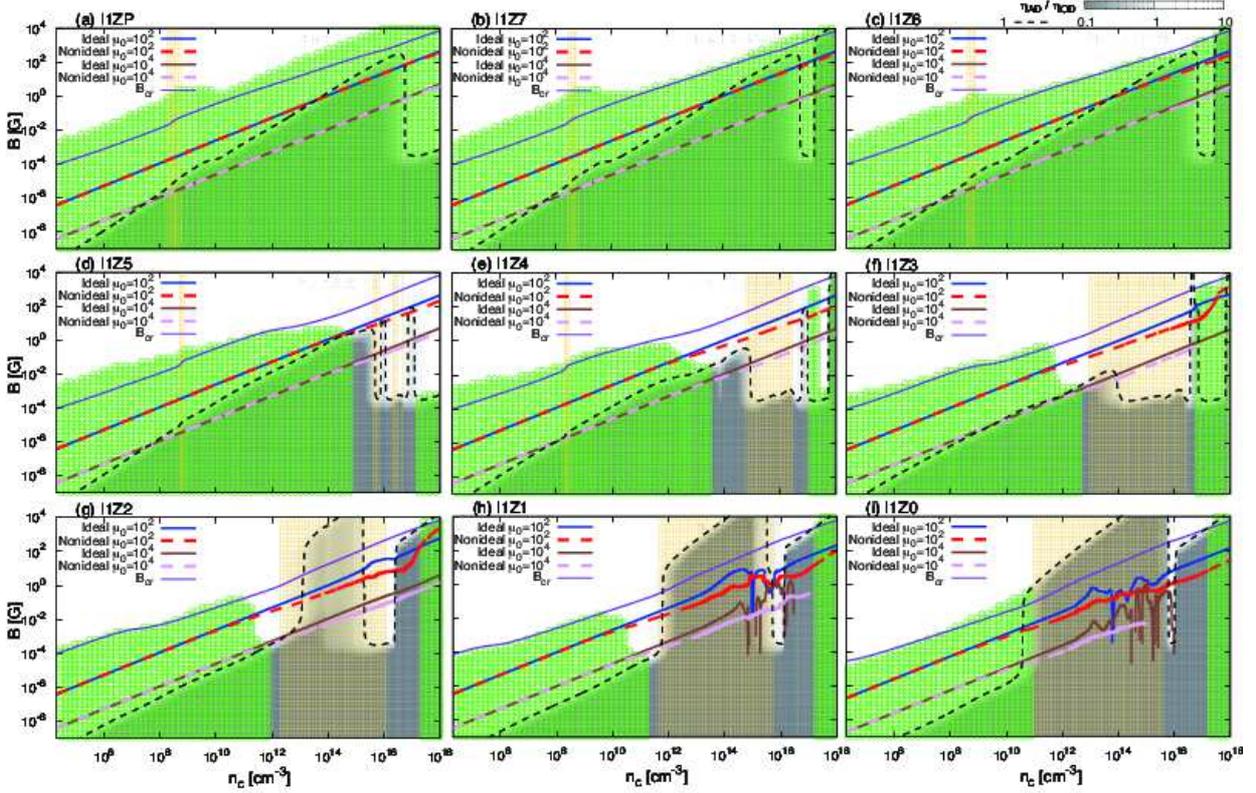}
\caption{
Same as Fig.~\ref{fig:color0}, but for models with $C_\zeta=1$.
}
\label{fig:color1}
\end{figure*}

Figs.~\ref{fig:color0} and \ref{fig:color1} show the evolution of
the central magnetic field strength as functions of central density
($n_{\rm c}$), for models with $C_\zeta=0$ and  $C_\zeta=1$, respectively. The nine panels in
each of the figures correspond to the different metallicities.
The blue and brown curves denote the results of ideal MHD simulations with
$\mu_0=10^2$ and $10^4$, while the red and pink curves correspond to the
non-ideal MHD runs.

In these figures, the magnetic Reynolds number is larger than unity in the
green region.
The magnetic Reynolds number is defined as 
\begin{eqnarray}
    {R}_{m} \equiv \vect{v}_{\rm f} \lambda_{\rm{J}} \eta^{-1},
\label{eq:reynolds} 
\end{eqnarray}
where $v_f \equiv [(4/3) \pi G \lambda_j^2 \rho_c]^{1/2} $ is the free-fall velocity and $\lambda_j \equiv (\pi c_s^2 / G \rho_c)^{1/2}$ is the Jeans wavelength \citep{machida07}.
In Eq.~(\ref{eq:reynolds}), the magnetic diffusivity $\eta$ is
estimated to be $\eta = \rm{max}(\eta_{\rm OD}, \eta_{\rm AD})$.
Note that $R_m$ is basically the same as the ratio of the free-fall velocity
to the drift velocity of the field lines with respect to the gas at the
Jeans scale, which was taken as the indicator of the magnetic field dissipation in one-zone
models \citep{nakano86,maki04,maki07,susa15}. Hence, the magnetic field
is coupled to the gas in the green regions ($R_m > 1$), and so dissipates only slightly from the cloud.
Note that eq. (\ref{eq:reynolds}) is a rough indicator of magnetic dissipation because the cloud collapse delays due to the Lorentz and pressure gradient forces and the dissipation of magnetic field can occur in the region of $R_{\rm m}>1$.

From the condition of the mass-to-flux ratio $\mu = 1$, we derived the critical magnetic field strength above which the 
cloud cannot collapse is defined as 
\begin{eqnarray}
B_{\rm cr} = \left( \frac{16 \pi^2 G \rho_{\rm c}^2 \lambda_{\rm J}^2 }{3}  \right)^{1/2},
\end{eqnarray}
as described in \citet{susa15}. 
Moreover, $B_{\rm cr}$ is indicated by the purple curves.

For reference, the region of polytropic index in which $\gamma>4/3$ is indicated by yellow hatched lines, in which the polytropic index $\gamma$ is defined as
\begin{eqnarray}
    \gamma = \frac{\rho_{\rm c}}{P_{\rm c}} \left(\frac{\rm{d} \it{P}}{\rm{d} \rho} \right)_{\rm c},
\end{eqnarray}
which is derived from one-zone calculation and thermal pressure $P$ taken from one-zone calculation is used in our three dimensional calculation (\S\ref{sec:method}).
A quasi-hydrostatic core (referred to as the first core) can be formed in the yellow region with $\gamma>4/3$,
\footnote{
We confirmed that the first core formation epoch for the model with $Z=\zsun$ and $C_\zeta=1$ (the present-day case) agrees well with that reported in \citet{larson69} and \citet{masunaga00}.     
}
where the contraction timescale of the cloud tends to become longer than the dissipation timescale of the magnetic field.
In addition, the ratio of ambipolar diffusivity $\eta_{\rm AD}$ to Ohmic resistivity $\eta_{\rm OD}$ is painted by the grey colour scale, and $\eta_{\rm AD}/\eta_{\rm OD}=1$ is plotted by the black
broken line, above which ambipolar diffusion dominates Ohmic dissipation
as the dissipation process of the magnetic field.

Now, we look into the details.
In Fig.~\ref{fig:color0}{\it a} (model I0ZP), i.e., the primordial case, we find that the evolution of the $B$-field in the ideal and non-ideal runs are almost identical for the
same $\mu$. This was described in a different manner in Fig.~\ref{fig:mu_each}a. In fact, the locus of the runs pass through the green
region, where ${\it R}_{m}>1$ is satisfied.
Moreover, the entire domain of the magnetically supercritical region (below
the purple line), is covered by the green region.
Hence, as described in \S\ref{sec:results}, the magnetic field rarely
dissipates in the primordial environment, which is well understood from
this figure. This result is fully consistent with the order-of-magnitude
arguments of one-zone models \citep{maki04,susa15}.
This high conductivity of the primordial gas is due to the presence of ionized
lithium (Li$^+$), which increase the ionization degree and maintain the
coupling between the magnetic field and neutrals \citep{maki04,susa15}. For the case in which dust grains exist, they absorb Li ions to reduce the
ionization degree \citep{maki04,susa15}.
The figure also indicates that ambipolar diffusion dominates Ohmic
dissipation for the models of $\mu_0=10^2$, while Ohmic
dissipation dominates ambipolar diffusion for $\mu_0=10^4$ calculation
models, although dissipation is not very important at the Jeans scale.

Figs.~\ref{fig:color0}{\it b} through \ref{fig:color0}{\it i} represent cases in which there is no ionization source ($C_\zeta = 0$) with various non-zero metallicities.
In these models, the magnetically
inactive region (area without green hatched lines) appears in the magnetically supercritical state (below the purple line). 
As a result, the evolutionary track of non-ideal MHD calculations
begins to deviate from that of ideal MHD calculations near the edge of the
magnetically active region (i.e., the boundary of the green region).
Thus, we can confirm that the dissipation of the magnetic field occurs in
the magnetically inactive region and that the order-of-magnitude arguments based on the value of $R_m$ are roughly valid. 
However, it is also true that, even in the magnetically inactive region,
$B$ gradually increases in non-ideal runs, although they  deviate from the
ideal MHD runs. Thus, we have to keep in mind that the magnetic field does have some
interactions with the gas, even in the region in which $R_m < 1$ is satisfied.

In Figs.~\ref{fig:color0}{\it b} through \ref{fig:color0}{\it i}, we can also see that the dominant dissipation process (Ohmic dissipation or ambipolar diffusion) depends on parameters $\mu$ and $Z/\zsun$. 
For example, in Fig.~\ref{fig:color0}{\it b}, Ohmic dissipation
acts as the main dissipation process of the magnetic field for the
$\mu_0=10^4$ calculation model, while both ambipolar and Ohmic
dissipation work for the models with $\mu_0=10^2$. 
Overall, however, for the $C_\zeta=0$ models (Fig.~\ref{fig:color0}), Ohmic dissipation is the dominant process of magnetic dissipation, although ambipolar diffusion dominates  Ohmic dissipation in a narrow density range for $\mu_0=10^2$  models I0Z7 (Fig.~\ref{fig:color0}{\it b}), I0Z6 (Fig.~\ref{fig:color0}{\it c}), I0Z5 (Fig.~\ref{fig:color0}{\it d}), and I0Z4 (Fig.~\ref{fig:color0}{\it e}). 

In Fig.~\ref{fig:color0}, the yellow regions (the density range with
$\gamma>4/3$) are found in models with $Z/\zsun \ge 10^{-5}$
(panels [{\it d}] through [{\it i}]). It is known that the adiabatic core forms  
when the model has a sufficiently wide range of density with $\gamma>4/3$ (or the wide yellow region in the figure, \citealt{machida09b}). 
Once the first core forms, it oscillates around an equilibrium state,
thereby the magnetic field $B$ also oscillates, as shown in Figs.~\ref{fig:color0}{\it f} through \ref{fig:color0}{\it i}. 

Fig.~\ref{fig:color1} shows the models with $C_\zeta=1$, which correspond to the nearby star-forming environment in our galaxy.
Models I1ZP (Fig.~\ref{fig:color1}{\it a}), I1Z7
(Fig.~\ref{fig:color1}{\it b}), and I1Z6 (Fig.~\ref{fig:color1}{\it c})
are totally magnetically active in the magnetically supercritical state
$B<B_{\rm cr}$ (below purple line), as in the primordial model I0ZP. 
Thus, in these models, the evolutionary track of $B$ in the ideal MHD calculations is in good agreement with that in the non-ideal MHD calculations. 

On the other hand, there is a noticeable difference in $Z/\zsun \ge10^{-5}$ models.
For model I1Z5M100 (Fig.~\ref{fig:color1}{\it d}), the magnetic field strength is $B=4.6 \times 10^2$\,G at $n_c \sim 10^{18}~{\rm cm^{-3}}$ in the ideal MHD calculation, while $B=2.2 \times 10^2$\,G in the non-ideal MHD calculation at the same epoch. 
As also seen in Fig.~\ref{fig:color1}{\it d}, for the $Z/\zsun=10^{-5}$ model, a dissipative region of magnetic field appears in the range of $10^{15}\cm \lesssim n_{\rm c} \lesssim 10^{17}\cm$, which causes a difference in magnetic field strength between ideal and non-ideal MHD calculations. 
The magnetically dissipative region also exists in Fig.~\ref{fig:color1}{\it e}--{\it i}, and thus the magnetic fields in non-ideal MHD calculations are always weaker than those in ideal MHD calculations for $Z/\zsun \ge 10^{-5}$ models. 
Thus, the ideal MHD assumption appears to hold for $Z/Z_\odot \lesssim 10^{-5}$ in the models of $C_{\zeta}=1$.
In models I1Z4 (Fig.~\ref{fig:color1}{\it e}), I1Z3
(Fig.~\ref{fig:color1}{\it f}), I1Z2 (Fig.~\ref{fig:color1}{\it g}),
I1Z1 (Fig.~\ref{fig:color1}{\it h}), and I1Z0 (Fig.~\ref{fig:color1}{\it
i}), both ambipolar diffusion and Ohmic loss act as the dissipation
process of the magnetic field, in which the dominant process depends on both the magnetic field strength and density. 
On the other hand, Ohmic dissipation always dominates ambipolar diffusion for model I1Z0 in the range of $B<B_{\rm cri}$, which was been reported in a number of previous studies \citep[e.g.,][]{nakano02}. 

\subsection{Adiabatic collapse phase and magnetic field dissipation}
Overlapping of the magnetically inactive regions and the adiabatic
region is quite important for the dynamics of collapsing prestellar core
\citep{machida08}.
In the present-day case, i.e., model I1Z0 (Fig. \ref{fig:color1}{\it i}),
the adiabatic phase (i.e., the first core phase) appears slightly earlier
than the time when the magnetic field becomes dissipative. As a result, the first core
is magnetically active when it is formed. Since the collapse of the core is slow in
this phase, the field lines are thought to have sufficient time to be
twisted in order to launch the outflow \citep{machida08,bate14}. 
As such, after the first core formation, various phenomena, such as the outflow driving, fragmentation, and spiral structure formation due to gravitational instability, which are closely related to the dissipation process of the magnetic field, appear \citep{machida08,machida14}. 
For lower metallicities and various $C_\zeta$s, we observed many interesting
overlap scenarios of the regions in Figs.~\ref{fig:color0}
and ~\ref{fig:color1}. This diversity should be important to the star
formation processes in various environments \citep{susa15}.
We will focus on this issue in a forthcoming paper. 
Note that these phenomena were not observed in the calculations in the present study because an initially weak magnetic fields and a slow rotation rate are adopted in order to focus only on the dissipation process of the magnetic field (\S\ref{sec:methods}). 

\subsection{Magnetic field during second-generation star formation }
As described in \S\ref{sec:results} and \S\ref{sec:nBplane}, metal enrichment increases the
resistive region. However, when ionization sources do not exist
(i.e., $C_\zeta=0$), even a trace amount of metals ($Z/\zsun \ge
10^{-7}$) can cause significant magnetic dissipation. This is because the dust
grains absorb charged particles, such as Li$^+$, to reduce the ionization degree of the
gas in the absence of non-thermal ionization processes \citep{susa15}.
Such low-ionization environments are expected to be realized just after the death of the first stars. 
From these environments, the second generation of stars is expected to
be born \citep{smith15,ritter15,chen17}, which are considered to be the HMP/EMP stars in our
galactic halo \citep[e.g.,][]{bc05}.
In this case, the present results indicate that the magnetic field might not play a significant role in
the star formation process because the magnetic field dissipates from the star-forming cloud. 
Since the magnetic field transfers angular momentum in the collapsing
cloud, the dissipation of the magnetic field and the rotation may cause fragmentation and the formation of binary and/or multiple stellar systems. 

\section{Summary}
\label{sec:summary}
The magnetic field plays a crucial role in present-day star formation and is related to various phenomena observed in nearby star-forming regions, such as molecular outflows, optical jets, fragmentation (or binary formation), and circumstellar disk formation. 
On the other hand, the role of the magnetic field in different star-forming environments, such as the early universe or a starburst galaxy, is not clear. 
This is because, in addition to the uncertainty concerning the magnetic field strength, the dissipation rate of the magnetic field had not been investigated for star-forming clouds in different environments. 
The magnetic diffusivity is determined by the ionization degree, which strongly depends on the star-forming environment (ionization strength and cloud metallicity). 
Recently, \citet{susa15} estimated the ionization degree and magnetic diffusivity in various environments, calculating an enormous number of chemical reactions with parameters of the ionization strength and metallicity. 
However, the evolution and dissipation rates of the magnetic field cannot be estimated directly from \citet{susa15}, because they used one-zone calculations with a parameter of the magnetic field strength.
In the present study, through three-dimensional non-ideal MHD simulations, we investigated the evolution and dissipation of the magnetic field in star-forming clouds embedded in various environments.
In the simulations, we used a table generated by one-zone calculations \citep{susa15}, in which coefficients of Ohmic dissipation and ambipolar diffusion are listed with respect to the density, temperature, and magnetic field strength.

We assumed 36 different environments with parameters of the ionization strength $C_\zeta$ (=0, 0.01, 1, 10) and metallicity $Z/\zsun$ (= 0, $10^{-7}$, $10^{-6}$, $10^{-5}$, $10^{-4}$, $10^{-3}$, $10^{-2}$, $10^{-1}$, 1).
Then, we prepared star-forming clouds in a nearly-equilibrium state with mass-to-flux ratios of $\mu_0=10^2$ and $10^4$ and calculated their evolution until the central density exceeds $\nc\gtrsim10^{18}\cm$ with and without magnetic dissipation. 
We performed 288 calculations in total and obtained the following results: 

\begin{itemize}
\item 
The magnetic field does not dissipate in purely
	  primordial environments ($C_\zeta=0$ and $Z/\zsun=0$), i.e., the
	  ideal MHD assumption is valid when we consider the dynamics at Jeans scale.
The magnetic field lines are also frozen in the gas, even in very metal poor environments, when the ionization source exists. For instance, if we assume $C_\zeta=1$, i.e., the strength at the level of ISM, the dynamics of gas clouds with $Z/\zsun < 10^{-5}$ can be regarded as ideal MHD.
Thus, in such environments, the magnetic field can play a significant role in the star formation process, unless the magnetic field is extremely weak.
However, for the case in which $C_\zeta=0$, a very small amount of metals
	  (dusts) makes the gas resistive. This kind of environment could be
	  realized at the site of second-generation star formation, where
	  the magnetic field could be less important than the purely
	  primordial case.
 \item As the ionization intensity increases, the coupling between magnetic field and neutrals recovers. 
Thus, the magnetic field plays a role in such environments. 
The dissipation rate of the magnetic field strongly depends on the
	   ionization strength, the  metallicity, and the magnetic field
	   strength, which greatly influence the dynamics of collapsing star-forming clouds.
	   Thus, the star formation process may considerably differ from place to place in a galaxy because the ionization strength should differ in each star-forming region. 
For example, the magnetic field can strongly influence the star formation process when the star-forming region is located in a massive star-forming region, in which the ionization intensity is strong. In addition, we observed a wide variety of patterns of overlap between the density range where the gas is resistive and that in which the cloud behaves adiabatically. This will lead to a wide variety of star formation mechanisms at different times and locations.  
\item Both ambipolar diffusion and Ohmic dissipation should be included in order
	  to correctly estimate the evolution of the magnetic field. 
The dominant process of the magnetic dissipation depends on the star-forming environment. 
When there are no ionization sources, Ohmic dissipation almost always dominates ambipolar diffusion.
However, as the ionization intensity becomes strong, the coefficient of Ohmic
	  dissipation decreases and ambipolar diffusion often dominates. When the ionization strength becomes large
	  enough ($C_\zeta>0.01$), ambipolar diffusion usually works as the
	  primary dissipation process of the magnetic field for low-metallicity
	  environments of $10^{-5} \le Z/\zsun < 10^{-2}$. 
For higher metallicities of $Z/\zsun \ge 10^{-2}$, Ohmic loss substantially dominates ambipolar diffusion. 
However, since the dominant process also depends on the magnetic field strength, we should include both processes in order to investigate the star formation in different environments.
\end{itemize}

\section*{Acknowledgements}
This study has benefited greatly from discussions with ~K. Tomida, ~K. Omukai, and ~M. Sekiya.
The present research used the computational resources of the HPCI system provided by (Cyber Sciencecenter, Tohoku University; Cybermedia Center, Osaka University, Earth Simulator, JAMSTEC) through the HPCI System Research Project (Project ID:hp150092, hp160079, hp170047).
The present study was supported by JSPS KAKENHI Grant Numbers
JP17K05387, JP17H02869, JP17H01101, and JP17H06360.
Simulations reported in this paper were also performed by 2017 Koubo Kadai on
Earth Simulator (NEC SX-ACE) at JAMSTEC.
This work was partly achieved through the use of supercomputer system SX-ACE at the Cybermedia Center, Osaka University.

\clearpage
\appendix
\section{Resolution Study}
\label{sec:appendix}
In this study, we investigated the dissipation of magnetic field in different environments using three-dimensional non-ideal MHD simulations and showed that the dissipation rate of magentic field strongly depends on the environment of each star forming cloud. 
In present calculations, we resolved the Jeans wavelength at least 16 cells.
However, the dissipation rate of the magentic field may change when the spatial resolution is changed. 
To confirm the dependence of the magnetic dissipation rate on the spatial resolution, we investigated the amplification of magnetic field for model I0Z4M100.
This model (I0Z4M100) has a large dissipation region of magnetic field.
In addition, we can calculate the cloud density until $n_{\rm c} < 10^{18}\cm$ as shown in Fig.~\ref{fig:color0}. 

\begin{figure*}
\includegraphics[scale=1.00]{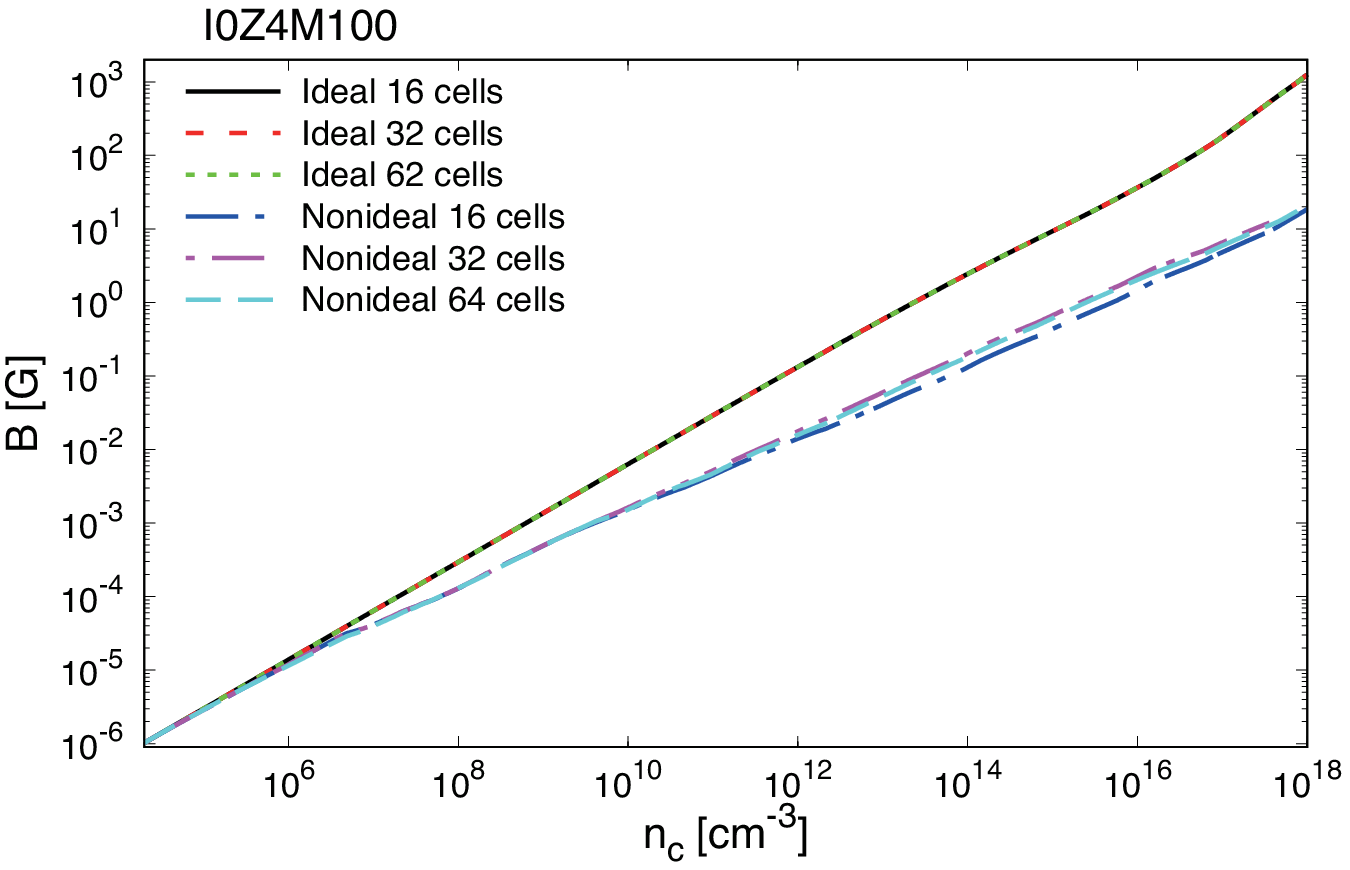}
\caption{
Magnetic field strengths $B$ at the centre for model I0Z4M100 with respect to the central number density ($n_{\rm c}$), in which  Ideal and Nonideal (MHD) models are plotted.
In each calculation, the Jeans wave length is resolved at least 16, 32, and 64 cells.
}
\label{fig:A1}
\end{figure*}

Fig.~\ref{fig:A1} plots the magentic field strengths for ideal and non-ideal (inclusion of Ohmic dissipation and ambipolar diffusion terms) MHD calculations against the central number density, in which the Jeans length is resolved at least 16, 32 and 64 cells. 
Thus, the spatial resolution differs among the calculations.
The figure indicates that, for ideal MHD models, the amplification of magnetic field does not significantly depend on the spatial resolution. 
On the other hand, we can confirm a small difference among non-ideal MHD models. 
The difference is seen  in the range of $10^{11}\cm \lesssim n_{\rm c} \lesssim 10^{18}\cm$, during which the magnetic dissipation is effective. 
The difference of magnetic field between non-ideal MHD models is about a factor of two at the maximum. 
Note that the magnetic field is smallest in Nonideal 32 cells models, but neither in 16 and 64 cells models. 
Note also that, thus, the high-spatial resolution is not necessary to give the most effective dissipation rate of magnetic field. 

Although the difference in magnetic field (a factor of two) may be large, the difference between ideal and non-ideal MHD 
models is about a factor of 100. 
Thus, we can safely estimate the dissipation of magnetic field with the resolution adopted in this study.

\end{document}